\documentclass[twocolumn,twocolappendix]{aastex631}

\received{June 9, 2023}
\revised{December 15, 2023}
\accepted{December 17, 2023}

\submitjournal{PSJ}

\shorttitle{Hydrated Minerals on TNOs}
\shortauthors{Seccull, Fraser, Kiersz, \& Puzia}
\graphicspath{{./}{figures/}}

\usepackage{soul,xcolor}

\begin{document}
\setstcolor{red}
\title{Hunting for Hydrated Minerals on Trans-Neptunian Objects}

\correspondingauthor{Tom Seccull}
\email{tom.seccull@proton.me}

\author[0000-0001-5605-1702]{Tom Seccull}
\affiliation{Gemini Observatory/NSF's NOIRLab, 670 N. A'ohoku Place, Hilo, HI 96720, USA}

\author[0000-0001-6680-6558]{Wesley C. Fraser}
\affiliation{Herzberg Institute of Astrophysics, 5071 West Saanich Road, Victoria, BC V9E 2E7, Canada}
\affiliation{Department of Physics and Astronomy, University of Victoria, Victoria, BC V8P 5C2, Canada}

\author[0000-0001-5787-9034]{Dominik A. Kiersz}
\affiliation{Independent Researcher, Belfast, UK}

\author[0000-0003-0350-7061]{Thomas H. Puzia}
\affiliation{Institute of Astrophysics, Pontificia Universidad Cat\'olica de Chile, Av. Vicu\~na MacKenna 4860, 7820436, Santiago, Chile}



\begin{abstract}
We present new optical reflectance spectra of three potentially silicate-rich trans-Neptunian objects (TNOs). These spectra were obtained with the aim of confirming past hints and detections of $\lambda\sim0.7~\mu$m absorption features associated with the presence of iron-bearing phyllosilicates. Our new spectrum of 120216~(2004~EW$_{95}$) presents clearly detected absorption features that are similar in shape to hydrated mineral absorption bands present in the spectra of aqueously altered outer main belt asteroids. Four new reflectance spectra of 208996~(2003~AZ$_{84}$) obtained at separate epochs all appear featureless, but they vary significantly in spectral gradient (between $\sim3.5\%/0.1~\mu$m and $\sim8.5\%/0.1~\mu$m) on a timescale consistent with this object's nominal rotational period. We report the first four optical reflectance spectra of 90568~(2004~GV$_{9}$), finding them all to be featureless but consistent with colors previously reported for this object. We speculate that impacts are the only mechanism capable of delivering, excavating, or forming hydrated minerals at the surfaces of TNOs in detectable concentrations; as a result, any deposits of hydrated minerals on TNOs are predicted to be localized and associated with impact sites. Globally altered TNOs (as observationally suggested for 2004~EW$_{95}$) plausibly formed more easily at smaller heliocentric distances ($r_H<15~$au) before being transplanted into the current trans-Neptunian population.       

\end{abstract}



\section{Introduction} \label{sec:intro}
Aqueous alteration is a process in which chemical alteration of anhydrous silicates is facilitated by liquid water molecules adsorbed to their surfaces and interfacial layers. These reactions produce hydrated minerals, which may include phyllosilicates, oxides, sulfides, and carbonates \citep[][and references therein]{2021GeCoA.299..219S}. Because aqueous alteration requires liquid water, there is a lower limit on the temperatures at which it occurs. So-called \textit{low-temperature} aqueous alteration occurs at temperatures above the freezing point of water ice ($273<T\lesssim400~$K), and is considered most relevant for the production of hydrated minerals observed both in primitive meteorites and on carbonaceous main belt asteroids \citep[e.g.,][]{2015aste.book..553W}. Hydrocryogenic alteration may occur in cases where thin interfacial layers of water remain unfrozen at the surfaces of silicates in ice-dust mixtures at $200<T<273~$K, and is potentially relevant to the production of hydrated minerals in comets \citep[e.g.,][]{1987ESASP.278..363R}. In cases where water forms a eutectic mixture with other volatiles like ammonia (NH\textsubscript{3}) and methanol (CH\textsubscript{3}OH), it may also potentially remain fluid down to $T\sim150$~K \citep[e.g.,][]{1992Icar..100..556K,2008ssbn.book..213M}. A detection of hydrated minerals on the surface of a minor planet is an extremely useful marker, indicating that at least some of its constituent silicates have been heated to temperatures at which aqueous alteration can happen. Such detections may assist constraint of not only the formation, thermal evolution, and dynamical evolution of individual minor planets but also that of the solar system more broadly \citep[e.g.,][]{2015aste.book...65R,2015aste.book..553W,2021GeCoA.299..219S}.

Unambiguous detection of hydrated minerals on the surfaces of main-belt asteroids and in aqueously altered meteorites is achieved through detection of a strong and distinctive $2.7-2.8~\mu$m absorption band in their reflectance spectra that is attributed to the fundamental stretch mode of structural hydroxyl \citep[OH; e.g.,][]{1980AJ.....85..573L,1990Icar...88..172J,1994Icar..111..456V,2005Natur.435...66C,2011epsc.conf..637H,2012Icar..219..641T,2015aste.book...65R,2022PSJ.....3..153R}. Absorption bands at 1.95, 2.95, and $4.3~\mu$m may be attributed to bound water within hydrated minerals, but these are not always easily distinguished from overlapping water ice absorption bands \citep[e.g.,][]{2005Natur.435...66C}. 

At near-ultraviolet (NUV) and optical wavelengths aqueously altered objects may show a variety of spectral behaviors. The most common of these are weak absorption features centered near $\lambda\sim0.7~\mu$m, and an absorption edge at $\lambda<0.55~\mu$m that results from the presence of multiple overlapping absorption bands centered at UV wavelengths. Features in both of these locations may be attributed to Fe\textsuperscript{2+}$\rightarrow$Fe\textsuperscript{3+} intervalence charge transfer (IVCT) transitions in oxidized iron present in phyllosilicates \citep[e.g.,][]{1989Sci...246..790V,1994Icar..111..456V,2011Icar..212..180C,2014Icar..229..196F,2014Icar..233..163F,2015aste.book...65R}. Ferric aqueous alteration features in the $0.7~\mu$m region are observed with a variety of shapes, central wavelengths, and substructures generally falling in the range $0.55-0.85~\mu$m, their precise configuration depending on the specific mineralogy of the surface. Bands centered at $0.7-0.75~\mu$m are typically attributed to serpentine group phyllosilicates, and those centered at $0.59-0.67~\mu$m are attributed to saponite group phyllosilicates \citep{2011Icar..212..180C}. Less frequently, ferric alteration bands have also been observed at slightly longer and shorter wavelengths, notably near $0.43~\mu$m and $0.8-0.9~\mu$m \citep[e.g.][]{1994Icar..109..274V}. Importantly, however, we note that the mineralogy of the materials absorbing in the $0.7~\mu$m region is not always straightforwardly deciphered by studying the varied and often complex forms of their absorption bands \citep[e.g.,][]{1989Sci...246..790V,2011Icar..212..180C}.

Most asteroid reflectance spectra with $0.7~\mu$m features also have a $2.7~\mu$m band, but around half of those with the $2.7~\mu$m band lack absorptions near $0.7~\mu$m \citep[][]{2011epsc.conf..637H}. This means that observation of a $0.7~\mu$m hydrated mineral feature is a good indication of the presence of hydrated minerals on the surface of a minor planet, but its absence does not necessarily indicate the opposite \citep{2015aste.book...65R}.

The presence of aqueously altered silicates in interplanetary dust particles and the dust of some comets \citep[e.g.,][]{2006Sci...313..635L,2022NatAs...6..731K} supports predictions that they may also be present on the surfaces of trans-Neptunian objects \citep[TNOs; e.g.,][]{2012ApJ...749...33F}. They should not necessarily be expected to be detectable in our remotely sensed reflectance spectra, however, especially at optical and near-infrared (NIR) wavelengths. This is because refractory organic residues are also likely to be present on the surfaces of TNOs and are predicted to govern their diverse optical-NIR colors \citep[e.g.,][]{1998Icar..135..389C,2005AdSpR..36..178C,2006A&A...455..725B,2006ApJ...644..646B,2012ApJ...749...33F,2013Icar..222..307D,2015Icar..252..311D,2017A&A...604A..86M,2017AJ....153..145W,2020Sci...367.3705G,2021PSJ.....2...10F}. Carbonaceous materials are known to mask the weak absorption bands of phyllosilicates, even when they are added to mixtures of silicates and ices in tiny concentrations \citep[e.g.,][]{2011Icar..212..180C,2016Icar..267..154P,2019GeoRL..4614307H}. Nevertheless, a handful of TNOs and centaurs have been reported to present absorption bands in their reflectance spectra that are associated with hydrated minerals \citep{2003AJ....125.1554L,2004A&A...416..791D,2004A&A...421..353F,2009A&A...508..457F,2008A&A...487..741A,2009A&A...501..777G,2018ApJ...855L..26S}. Observations of hydrated mineral bands in the reflectance spectrum of a TNO therefore suggest that it has an unusually high abundance of surface hydrated minerals relative to its abundance of refractory organics.

With this in mind, this work is not concerned directly with the question of whether hydrated minerals are present on the surfaces of TNOs \citep[in many cases the answer appears likely to be yes;][]{2005Natur.435...66C,2022NatAs...6..731K}. Rather, when interpreting TNO reflectance spectra it may be more constructive to ask why and how some TNOs become \textit{observably} aqueously altered, why others do not, and whether the distinction between these two groups reveals anything about their respective pathways of formation and thermal evolution. Reliable and repeatable detection of aqueously altered TNOs through photometric observation and spectroscopic followup is a crucial first step toward untangling these questions, and our efforts to achieve this are the primary subject of this work. In Section \ref{sec:targ} we describe the methods we use to identify aqueously altered TNOs for followup spectroscopy. Section \ref{sec:obsred} presents the observation and reduction of new reflectance spectra of three TNOs that are predicted to be aqueously altered. Section \ref{sec:res} presents our analysis of these spectra. Finally, in Section \ref{sec:dis} we discuss our findings and make predictions about how aqueously altered material may become detectable on the surfaces of TNOs.   

\section{Target Selection} \label{sec:targ}
Our target TNOs were selected because they met three of four criteria, the first being mandatory:

\begin{enumerate}
\item Membership of the less red TNO color class \citep[$V-R<0.56$; ][]{2019AJ....157...94M}. TNOs with redder surfaces may be more likely to have their surface silicates masked, potentially due to a higher abundance of refractory organics \citep[e.g.,][]{2016Icar..267..154P}. Curvature in the spectra of these ultrared hydrocarbons may also be difficult to distinguish from curvature caused by weak bands near $\lambda\sim0.7~\mu$m \citep[e.g.,][]{2004Icar..168..158R}.
\item Reported infrared photometric colors (at $2.2<\lambda<5.0~\mu$m) that suggest a silicate-rich composition \citep{2021PSJ.....2...10F}.
\item Previously reported observations of a $0.7~\mu$m absorption band in their reflectance spectra \citep[e.g.,][]{2009A&A...508..457F}.
\item Published optical photometric colors in \citet{2015A&A...577A..35P} that hint at the presence of absorptions in their spectra near $0.7~\mu$m. Our method of testing the colors of TNOs is described in Appendix \ref{sec:colors}.
\end{enumerate} 

Based on these criteria, we chose to observe 90568 (2004~GV$_{9}$), 120216 (2004~EW$_{95}$), and 208996 (2003~AZ$_{84}$).        

\section{Observations \& Data Reduction} \label{sec:obsred}
Each of our targets was observed in longslit spectroscopy mode at Gemini Observatory with, depending on its declination, either the northern or southern Gemini Multi-Object Spectrograph \citep[GMOS;][]{2004PASP..116..425H,2016SPIE.9908E..2SG}. Alongside each TNO, and under identical observing conditions, we observed the spectrum of at least one solar twin selected from published catalogs \citep{2014A&A...563A..52P,2014A&A...572A..48R}. These calibrator star spectra were used to negate the solar signature in our TNO spectra and derive their reflectance spectra. A log of our observations, their geometry, and the instrument configurations used to obtain the data are presented in Appendix \ref{sec:supp}. The spectra of all our targets and calibrator stars were observed under photometric conditions (CC50 as defined by Gemini Observatory); as a result, we expect that the shapes and gradients of our reflectance spectra are accurate. 

For all our observations we used a repeating four point dither pattern (0\arcsec, +16\arcsec, +8\arcsec, -8\arcsec) along the slit to enable construction of fringe frames and aid mitigation of bad pixels during stacking. During the acquisition of each target, the slit was aligned to the average parallactic angle calculated to occur during the associated sequence of spectroscopic integrations. We used the standard GMOS Hamamatsu CCD detectors and binned our spectra $2\times2$ (spatial~$\times$~spectral) on the detector before reading out the central spectrum region of interest in the standard slow read, low gain, 12 amplifier mode. Once at the start of each spectroscopic sequence targeting a TNO, and once again after each cycle of four dithers, a pair of spectroscopic lamp flats were observed. Five bias frames were obtained for each TNO observation during the standard sequence of morning calibrations. A single arc frame was also collected by observing a CuAr calibration lamp during daytime calibrations following each observation of 2003~AZ$_{84}$ and 2004~GV$_{9}$. As a precaution against the effects of instrument flexure, the arc lamp frame used to calibrate the spectrum of 2004~EW$_{95}$ was obtained while on sky, in the middle of the science sequence. The effects of flexure on the wavelength calibration of GMOS spectra observed in the red optical region are minimal, however, and we did not notice any adverse effects in our data caused by use of arc frames obtained during daytime.

On 2022 January 8, an uncontrolled warmup of GMOS-S effectively disabled amplifier 5 of the GMOS-S detector for science purposes.\footnote{\url{https://www.gemini.edu/news/instrument-announcements/gmos-s-ccd2-issues}} Attempts to adapt our observing strategy for our initial observation of 2004~GV$_{9}$ were unfortunately unsuccessful and caused the center of that spectrum to fall on the area of the detector read out by the damaged amplifier. As a result, data in the range $0.66-0.78~\mu$m were unrecoverable; there is now a gap at these wavelengths in our reflectance spectra of 2004~GV$_{9}$ obtained at the first epoch on 2022 March 7.

Initial data reduction steps for all our data, including bias subtraction, flat-field correction, wavelength calibration, and 2D spectrum rectification, were performed with Gemini IRAF \citep{1986SPIE..627..733T,1993ASPC...52..173T,2012SASS...31..159G}. The solar calibrator star spectra were calibrated with the same biases, flats, and arcs as were used to calibrate their associated TNO spectra. Astroscrappy, a Python implementation of LACosmic \citep{2001PASP..113.1420V, 2018zndo...1482019M}, was then used to remove cosmic rays from the 2D spectra. For each 2D frame observed at a given spatial dither point, we subtracted a fringe frame created by median combining the frames at all other dither points. This aimed to correct minor fringing artifacts and any scattered light present in the background of the spectra not already flattened during flat-fielding.

Sky subtraction and spectrum extraction were then performed with our Modular Optimized Tracer and Extractor of Spectra (MOTES; T.~Seccull \& D.~A.~Kiersz 2024, in preparation). This software builds on the methods described by \citet{2018ApJ...855L..26S} used for localization and extraction of faint point-source spectra in that it employs a version of optimal extraction \citep{1986PASP...98..609H}. MOTES returns both an optimally extracted spectrum and an aperture extracted spectrum for each 2D spectrum provided to it; a brief overview of the operation of MOTES is provided in Appendix {\ref{sec:motes}}. Because of the large gap in our first-epoch spectra of 2004~GV$_{9}$, we split each one in half to sky subtract and extract each good section separately.

As any flux calibration of our data would be canceled out during division by the solar calibrator spectra, and since the responses of the GMOS spectrographs are stable at least over the time span required to complete one of our observations ($\lesssim2~$hr), we elected not to perform a flux calibration on our extracted spectra. We did, however, apply a consistent relative extinction correction to each 1D spectrum (including both TNOs and calibrators) with $f{_C}(\lambda) = f(\lambda)10^{0.4ak(\lambda)}$, where $f(\lambda)$ is the uncorrected spectrum, $f_{C}(\lambda)$ is the corrected one, $a$ is the median airmass during the integration of the spectrum, and $k(\lambda)$ is the optical extinction curve stored in Gemini IRAF after it has been interpolated to the resolution of the data \citep[see][ and the Gemini Observatory website\footnote{\url{https://www.gemini.edu/observing/telescopes-and-sites/sites}}]{1983MNRAS.204..347S,1984MNRAS.206..241B}.         

Following extinction correction, all extinction-corrected 1D spectra for each TNO and calibrator star from each observation were median stacked. The TNO spectra were then divided by the spectra of one of their associated solar calibrator stars to produce reflectance spectra of the TNOs. These reflectance spectra were then binned using the bootstrapping method described by \citet{2019AJ....157...88S} to boost their signal-to-noise ratio (S/N). For 2004~GV$_{9}$, 2004~EW$_{95}$, and 2003~AZ$_{84}$ we used binning factors (i.e., numbers of points per bin) of 12, 30, and 66, respectively. We note, however, that higher binning factors were used in cases where our data had poor quality. Our third 2003~AZ$_{84}$ reflectance spectrum has a binning factor of 94, because although this sequence of spectroscopic observations (observed 2022 Jan 23) was obtained with all required calibrations, it was only partially completed. Our first epoch reflectance spectra for 2004~GV$_{9}$ have a binning factor of 60. Our method of selecting binning factors is detailed in Appendix \ref{sec:binning}.

The lower wavelength limit of our 2004~EW$_{95}$ reflectance spectrum is defined by the cutoff of the OG515 order blocking filter. The other two spectra had both their shortest and longest wavelength extremities removed. The efficiencies of both the R150 and R400 gratings drop rapidly at $\lambda\lesssim0.6~\mu$m\footnote{\url{http://www.gemini.edu/instrumentation/gmos/components}} along with delivered S/N. In all cases we cut out data at $\lambda\gtrsim0.90~\mu$m owing to the presence of large residuals caused by incomplete cancellation of sky emission lines and telluric absorption lines. The spectra of 2004~GV$_{9}$ and 2003~AZ$_{84}$ also suffered second-order contamination at $\lambda>0.91~\mu$m.

\section{Results \& Analysis} \label{sec:res}
We present our new reflectance spectra of 2004~GV$_{9}$ and 2004~EW$_{95}$ in Figures \ref{fig:gv9} and \ref{fig:ew95}, respectively. Figure \ref{fig:ewmodel} presents the spectrum of 2004~EW$_{95}$ reported by \citet{2018ApJ...855L..26S} in the context of colors reported for the same object by \citet{2015ApJ...804...31F} and the model of the shape of the spectrum reported by \citet{2020PhDT........S}. Figure \ref{fig:az84} presents our new reflectance spectra of 2003~AZ$_{84}$. If readily available, we compare published reflectance spectra of our targets to our own. We also compare our data to coarse reflectance spectra of our targets derived from published single-epoch colors observed both from the ground in the $VRI$ filters \citep{2004A&A...421..353F,2008AJ....136.1502R,2009A&A...494..693S,2010A&A...510A..53P,2013A&A...554A..49P,2016AJ....152..210T} and with Wide Field Camera 3 on the Hubble Space Telescope (HST) in the F606W, F775W, F814W, and F098M filters \citep{2015ApJ...804...31F}. Both $VRI$ and HST colors were converted to reflectance with methods described by \citet{2002A&A...389..641H}, $VRI$ solar colors reported by \citet{2012ApJ...752....5R}, and HST solar colors provided privately by the authors of \citet{2015ApJ...804...31F}. 

For purposes of comparison to literature values, and to quantify any detectable color variability, we measured the gradient of each of our new reflectance spectra ($S'$; see Table \ref{tab:grads}). For 2004~EW$_{95}$ and 2003~AZ$_{84}$ we measured the gradient across the full available wavelength range. This was not possible for the first-epoch reflectance spectra of 2004~GV$_{9}$ on account of the gap in their wavelength coverage, so we measured the gradients of all our 2004~GV$_{9}$ spectra in two sections. The first section covered the maximum available reliable spectroscopic data in the range $0.50-0.65~\mu$m; $0.56-0.66~\mu$m in the first-epoch spectra and $0.50-0.65~\mu$m in spectra from epochs 2, 3, and 4. The second section covered the maximum available reliable spectroscopic data in the range $0.65-0.90~\mu$m; $0.79-0.90~\mu$m in the first-epoch spectra and $0.65-0.90~\mu$m in spectra from epochs 2, 3, and 4. The dividing line was set at $0.65~\mu$m as the spectra of 2004~GV$_{9}$ appear consistent across observing epochs at $\lambda<0.65~\mu$m, but have slightly more varied gradients at $\lambda>0.65~\mu$m (see section \ref{sec:gv}). Our method of measuring spectral gradients is described in the Appendix \ref{sec:grads}. We now describe our findings for each of our targets.    

\begin{table}
\begin{center}
\caption{Spectral Gradients in $\%/0.1~\mu$m}
\label{tab:grads}
\begin{tabular}{p{3.7cm}p{0.7cm}p{0.7cm}p{2.1cm}}
\hline\hline
Target/Calibrator~\textit{(Epoch)} & $S'$ & $\delta{S'}$ & $\lambda$~Range ($\mu$m)\textsuperscript{a} \\[1pt]
\hline
2004~EW$_{95}$/HD 124523 & 3.6 & 0.1 & 0.54-0.91 \\[1pt]
\hline\hline
2003~AZ$_{84}$/HD~54351 \textit{(1)} & 3.7 & 0.9 & 0.58-0.92 \\[1pt]
2003~AZ$_{84}$/HD~78534 \textit{(1)} & 3.5 & 0.8 & 0.58-0.92 \\[1pt]
2003~AZ$_{84}$/HD~54351 \textit{(2)} & 9.1 & 1.7 & 0.58-0.92 \\[1pt]
2003~AZ$_{84}$/HD~78534 \textit{(2)} & 8.6 & 1.7 & 0.58-0.92 \\[1pt]
2003~AZ$_{84}$/HD~54351 \textit{(3)} & 3.8 & 1.4 & 0.58-0.92 \\[1pt]
2003~AZ$_{84}$/HD~78534 \textit{(3)} & 3.0 & 1.2 & 0.58-0.92 \\[1pt]
2003~AZ$_{84}$/HD~54351 \textit{(4)} & 8.4 & 1.1 & 0.58-0.92 \\[1pt]
2003~AZ$_{84}$/HD~78534 \textit{(4)} & 7.9 & 0.8 & 0.58-0.92 \\[1pt]
\hline\hline
2004~GV$_{9}$/HD~124523 \textit{(1)} & 20.0 & 1.3 & 0.56-0.66 \\[1pt]
2004~GV$_{9}$/HD~157750 \textit{(1)} & 21.0 & 1.2 & 0.56-0.66 \\[1pt]
2004~GV$_{9}$/HD~124523 \textit{(2)} & 19.6 & 0.4 & 0.50-0.65 \\[1pt]
2004~GV$_{9}$/HD~124523 \textit{(3)} & 19.6 & 1.8 & 0.50-0.65 \\[1pt]
2004~GV$_{9}$/HD~124523 \textit{(4)} & 20.2 & 0.4 & 0.50-0.65 \\[1pt]
2004~GV$_{9}$/HD~124523 \textit{(1)} & 10.3 & 3.1 & 0.79-0.90 \\[1pt]
2004~GV$_{9}$/HD~157750 \textit{(1)} & 10.6 & 3.1 & 0.79-0.90 \\[1pt]
2004~GV$_{9}$/HD~124523 \textit{(2)} & 10.1 & 0.2 & 0.65-0.90 \\[1pt]
2004~GV$_{9}$/HD~124523 \textit{(3)} & 12.6 & 0.4 & 0.65-0.90 \\[1pt]
2004~GV$_{9}$/HD~124523 \textit{(4)} & 11.4 & 0.4 & 0.65-0.90 \\[1pt]
\hline
\hline
\end{tabular}\\[2pt]
\end{center}
\small{\textbf{Note.} This table shows the spectral gradients ($S'$) and associated uncertainties measured from our spectra across various wavelength intervals.\\\textsuperscript{a} The rightmost column notes the wavelength interval over which each $S'$ measurement was made; all $S'$ values are measured relative to the reflectance at the center of this range.} \\
\end{table}

\subsection{90568 (2004~GV$_{9}$)}\label{sec:gv}
\begin{figure*}
\centering
\includegraphics[scale=0.64]{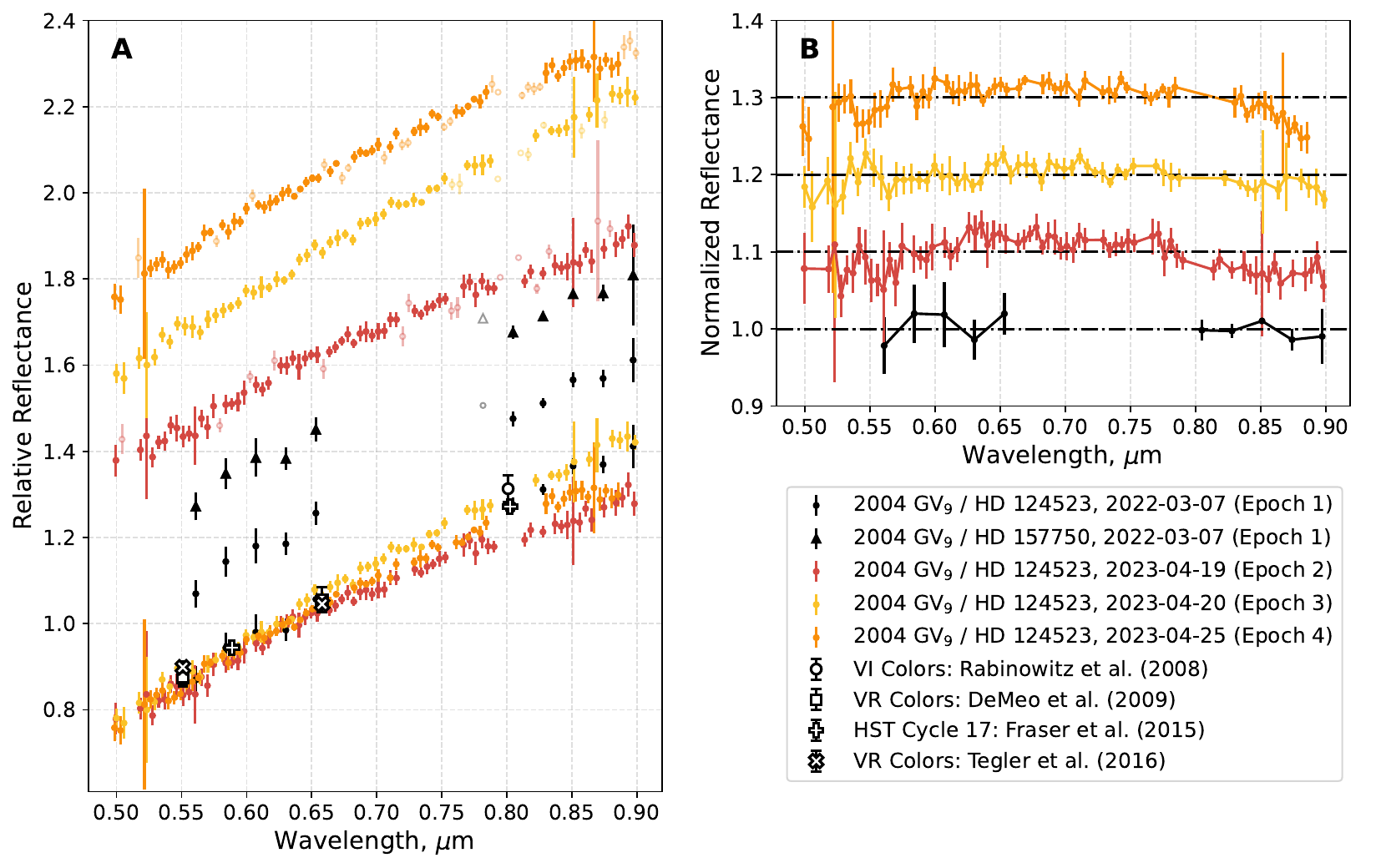}
\caption{Optical reflectance spectra of 2004~GV$_{9}$. \textbf{(A)}~Here we compare our new GMOS reflectance spectra of 2004~GV$_{9}$ both to each other and to coarse spectra derived from published single-epoch photometry \citep{2008AJ....136.1502R,2009A&A...493..283D,2015ApJ...804...31F,2016AJ....152..210T}. Hollow points in the GMOS spectra are affected by incomplete background subtraction, telluric band residuals, proximity to the GMOS chip gaps, or a combination of these effects. All datasets are scaled to unit reflectance at $0.625~\mu$m. We present all spectra calibrated with HD~124523 overlaid for direct comparison, and also offset for clarity. All offset spectra are offset vertically in increments of +0.2.  \textbf{(B)}~Here we show our 2004~GV$_{9}$ spectra following division by a line fitted to them across their full wavelength coverage. Spectra from each epoch are offset vertically for clarity in increments of +0.1. Only spectra calibrated with HD~124523 are shown. Points affected by residuals of telluric absorption or imperfect background subtraction have been omitted. \label{fig:gv9}}
\end{figure*}

2004~GV$_{9}$ is a moderately large hot classical TNO (see Appendix \ref{sec:supp}) that is predicted to have a silicate-rich composition by \citet{2021PSJ.....2...10F}. Its reflectance spectrum is flat and featureless at $1.4-2.4~\mu$m, indicating an absence of detectable surface ices \citep{2009Icar..201..272G,2012AJ....143..146B}. We could not find a full set of single-epoch $VRI$ colors in the literature for 2004~GV$_{9}$, but $V-R$, $R-I$, and F606W-F814W colors are available \citep{2008AJ....136.1502R,2009A&A...493..283D,2015ApJ...804...31F,2016AJ....152..210T}. In combination these colors suggest that the reflectance spectrum of 2004~GV$_{9}$ is linear and featureless at $0.55-0.8~\mu$m and does not significantly vary (see Figure \ref{fig:gv9}). When, however, we applied the photometric test for a $0.7~\mu$m band described in Appendix \ref{sec:colors} to the combined $VRI$ colors of 2004~GV$_{9}$ compiled by \citet{2015A&A...577A..35P} from \citet{2008AJ....136.1502R} and \citet{2016AJ....152..210T}, we found that they are compatible with the presence of a hydrated mineral absorption band.

To our knowledge, our reflectance spectra of 2004~GV$_{9}$ are the first to be reported covering optical wavelengths (see Figure \ref{fig:gv9}). They are all featureless, lacking anything but red continuum near $0.7~\mu$m, and appear consistent with coarse reflectance data derived from photometric colors previously reported for 2004~GV$_{9}$ \citep[Fig. \ref{fig:gv9};][]{2008AJ....136.1502R,2009A&A...493..283D,2015ApJ...804...31F,2016AJ....152..210T}. The gradients of all the spectra at $\lambda<0.65~\mu$m remain consistently at values of $S'=20{\pm}1~\%/0.1~\mu$m across all four observing epochs (see Table \ref{tab:grads}). Variation of the spectral continuum appears to increase at $\lambda>0.65~\mu$m. 

Mapping our spectroscopic sampling of the surface to the rotational phase of 2004~GV$_9$ presents a conundrum. The best sampled light curve published so far for 2004~GV$_9$ was reported by \citet{2008A&A...490..829D} alongside an estimated rotational period of $5.86\pm0.03$~hrs. Comparison of this period to our observing log suggests that we observed 2004~GV$_9$ at roughly the same rotational phase during epochs 2, 3, and 4. In this context the subtle variation of the spectral gradient observed at $\lambda>0.65~\mu$m seems puzzling. Our consistent methods of observation allow us to rule them out as potential causes of the color variation. Visual inspection of through-slit acquisition frames for both 2004~GV$_{9}$ and HD~124523 shows them to be well centered at each epoch. Likewise, the slit position angle was correctly aligned to the parallactic angle for all observations; this and the low airmass at which all these observations were performed should minimize wavelength-dependent slit losses due to atmospheric differential refraction \citep{1982PASP...94..715F}. Reduction of the spectra observed at all epochs was done consistently, and sky emission at $\lambda>0.65~\mu$m appears to have been subtracted equally well from the spectra at each epoch. The redness of 2004~GV$_{9}$ does not correlate with either the airmass at which it was observed or the airmass difference between 2004~GV$_{9}$ and HD~124523 at each epoch. Nor does redness appear to correlate with the delivered image quality (IQ). Low level variations in atmospheric transmission may occur on timescales of minutes at $\lambda>0.65~\mu$m due to variations in the strength of telluric water absorption bands \citep[see][]{2015A&A...576A..77S} caused by changes to the precipitable water vapor (PWV) present in Earth's atmosphere along an observer's line of sight. Although variation in PWV could be a convenient explanation for the observed variation in the spectrum of 2004~GV$_{9}$, it is not monitored by Gemini South, precluding checks for causal links between PWV and the redness of 2004~GV$_{9}$ at $\lambda>0.65~\mu$m. 

While we cannot rule out atmospheric conditions as a cause of the variation in our spectra, we equally cannot yet completely rule out the possibility that the color variation is intrinsic to the surface of 2004~GV$_{9}$, and that the uncertainty of the rotational period reported by \citet{2008A&A...490..829D} is underestimated. A discrepancy exists between the light curves published for 2004~GV$_{9}$ by \citet{2008A&A...490..829D} and \citet{2007AJ....134..787S}. \citet{2008A&A...490..829D} report a light curve amplitude of $0.16\pm0.03$ mag for 2004~GV$_{9}$. \citet{2007AJ....134..787S} reports a more sparsely sampled light curve of equal photometric quality covering a much longer total observational baseline, and concludes that there is no evidence of a short-term lightcurve for 2004~GV$_{9}$ with amplitude greater than 0.1 mag.  These light curves were observed only two years apart, representing a maximum possible change in viewing aspect along the rotational axis of $\sim2^{\circ}$; the projection of 2004~GV$_{9}$ should not have meaningfully changed between these two observational epochs. Even though the light curve sampling of \citet{2007AJ....134..787S} is sparser than that of \citet{2008A&A...490..829D}, the data have sufficient spacing and S/N to have been able to detect variations in brightness with the period and strength of those reported by \citet{2008A&A...490..829D}. Based on the discrepancy between these two light curves it remains difficult to conclusively rule out the possibility that the uncertainty of the rotational period reported by \citet{2008A&A...490..829D} is underestimated. As a result, it remains possible that we may have observed 2004~GV$_{9}$ at multiple rotational phases. In any case, our understanding of both the reflectance spectrum and lightcurve of 2004~GV$_{9}$ could clearly benefit from additional observational monitoring. 

We take the null detection of any $0.7~\mu$m hydrated mineral features as a demonstration that photometric tests for such bands \textit{must} be performed on color data observed at a single epoch to minimize the influence of rotational color variation on their results. Testing of averaged color data for hydrated mineral bands is not recommended.

\subsection{120216 (2004~EW$_{95}$)}\label{sec:ew}
\begin{figure*}
\centering
\includegraphics[scale=0.68]{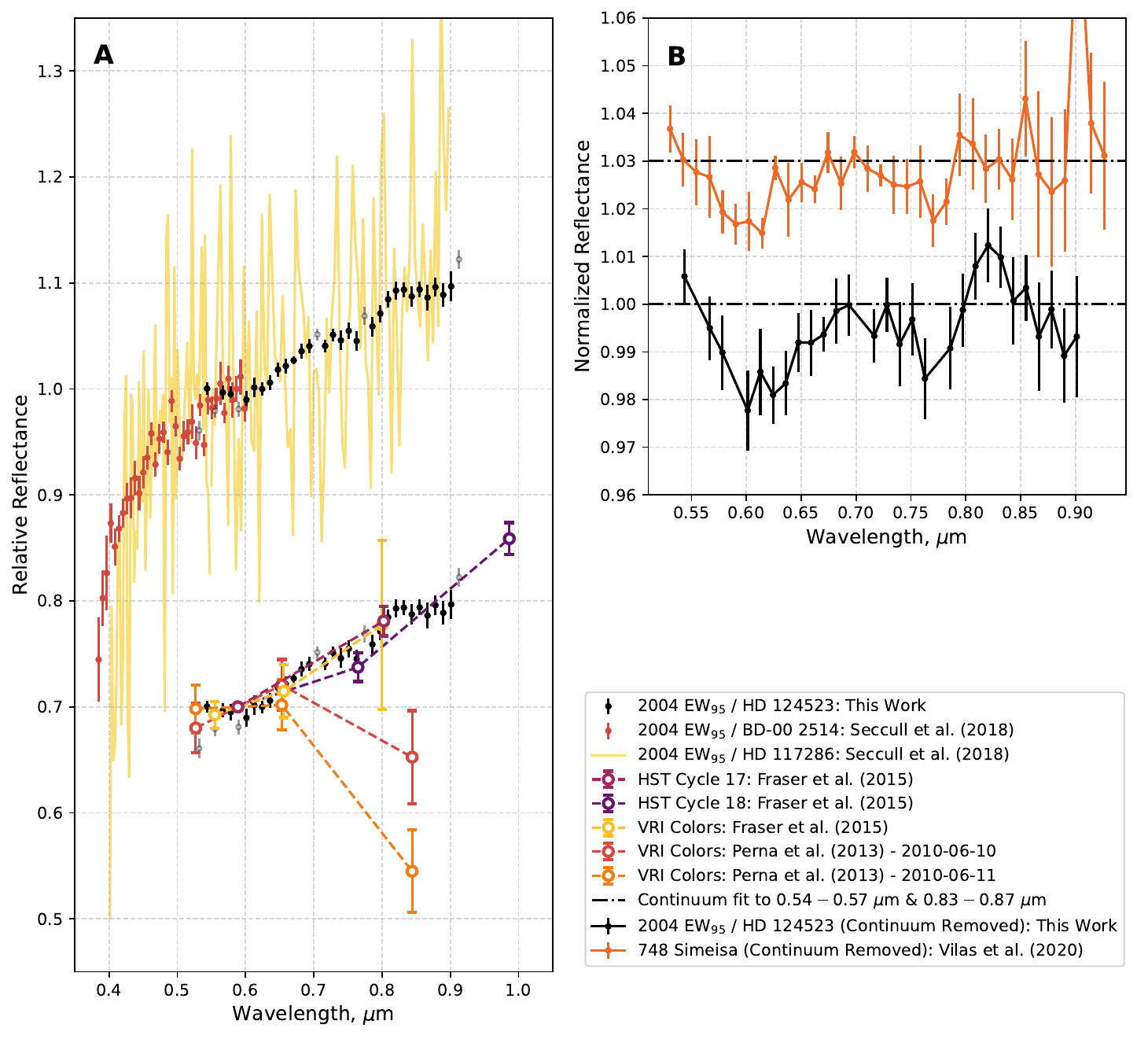}
\caption{The optical reflectance properties of 2004~EW$_{95}$. \textbf{(A)}~Our new GMOS reflectance spectrum of 2004~EW$_{95}$ is compared to those reported by \citet{2018ApJ...855L..26S}. The 2004~EW$_{95}$~/~BD-00~2514 spectrum presented here is the same as that reported by \citet{2018ApJ...855L..26S}, except that we have doubled its binning factor from 20 to 40. Coarse reflectance spectra produced from single-epoch colors are also shown \citep{2013A&A...554A..49P,2015ApJ...804...31F}. Note that the \citet{2015ApJ...804...31F} $VRI$ colors presented here are not the average colors reported in that work; instead, they are those observed in support of the \textit{TNOs are Cool} program \citep{2013A&A...554A..49P} and then reduced and published by \citet[][; $V-R=0.38\pm0.04$, $R-I=0.41\pm0.07$]{2015ApJ...804...31F}. All datasets are scaled to unit reflectance at $0.589~\mu$m. A copy of our GMOS spectrum and the coarse spectra overlaying it are similarly vertically offset by -0.3 for clarity. Hollow points in the GMOS spectrum are affected by incomplete background subtraction, telluric band residuals, proximity to the GMOS chip gaps, or a combination of these effects. \textbf{(B)}~Our GMOS spectrum of 2004~EW$_{95}$ after removal of the linear continuum (fitted at $0.54-0.57~\mu$m and $0.83-0.87~\mu$m). Points affected by residuals of telluric absorption or imperfect background subtraction have been omitted. For comparison we plot a reflectance spectrum of the Hilda asteroid 748 Simeisa that has been similarly continuum corrected, vertically offset by +0.03 for clarity, and binned to approximately the same resolution as that of our 2004~EW$_{95}$ spectrum. The 748 Simeisa spectrum was retrieved from the PDS Small Bodies Node \citep{2020PDS4..urn:nasa:pds:gbo.ast.vilas.spectra::1.0V}.\label{fig:ew95}}
\end{figure*}

2004~EW$_{95}$ is a relatively small 3:2 resonant TNO (see Appendix \ref{sec:supp}) that occupies a high inclination orbit undergoing high amplitude Lidov-Kozai oscillations \citep{2007Icar..189..213L}. It is predicted to be a silicate-rich object based on its infrared colors \citep{2021PSJ.....2...10F}. The reflectance spectrum of 2004~EW$_{95}$ drops precipitously from $\sim0.55~\mu$m toward NUV wavelengths, and it is also reported to have a $0.7~\mu$m hydrated mineral band \citep{2018ApJ...855L..26S}. The $VRI$ colors of 2004~EW$_{95}$ reported by \citet{2015ApJ...804...31F} imply the potential presence of a hydrated mineral band when tested as described in Appendix \ref{sec:colors}. Both the NUV reflectance drop and the $0.7~\mu$m band are consistent with those observed in the spectra of aqueously altered main-belt carbonaceous asteroids \citep{2014Icar..233..163F} and may be attributed to IVCT transition absorptions occuring in iron-bearing phyllosilicates \citep[e.g.,][]{1994Icar..109..274V,2015aste.book...65R}. In combination, the dynamical properties of 2004~EW$_{95}$ and its apparent compositional similarity to outer main belt asteroids have led to the prediction that it formed near to Jupiter and Saturn \citep{2018ApJ...855L..26S} before being scattered into the outer Solar System as those planets grew and migrated \citep[e.g.,][]{2012M&PS...47.1941W,2017Icar..297..134R,2020tnss.book...25M}. \citet{2013AJ....146....6L} showed that highly inclined, mean-motion resonant, Kozai resonant orbits, like that occupied by 2004~EW$_{95}$, are indeed more efficiently populated through resonance capture of scattering objects \citep[e.g.,][]{2008Icar..196..258L} than they are through a mechanism like resonance sweeping \citep[e.g.,][]{2005AJ....130.2392H}.

Our new reflectance spectrum of 2004~EW$_{95}$ (2004~EW$_{95}$/HD~124523) is presented in Figure \ref{fig:ew95}. The gradient of the spectrum measured across its full wavelength coverage ($0.54-0.91~\mu$m; see Table \ref{tab:grads}) is consistent with those reported previously at similar wavelengths \citep{2015A&A...577A..35P,2018ApJ...855L..26S}.  Removal of a linear continuum fitted at $0.53-0.57~\mu$m and $0.83-0.87~\mu$m reveals the presence of two weak absorption features that together are consistent in shape with those reported in the reflectance spectra of aqueously altered outer main-belt asteroids; for comparison, in Figure \ref{fig:ew95} we plot the continuum-removed spectrum of 2004~EW$_{95}$ against that of an aqueously altered Hilda asteroid, 748 Simeisa \citep{1989Sci...246..790V,2020PDS4..urn:nasa:pds:gbo.ast.vilas.spectra::1.0V}. For the stronger of the two bands, centered near $0.62~\mu$m, we measure an average depth of $1.8\pm0.4\%$ in the range $0.60-0.64~\mu$m relative to the fitted continuum. The location of this band suggests the presence of iron-bearing saponite group phyllosilicates, which typically have band centers in the range $0.59-0.67~\mu$m \citep[e.g.,][]{2011Icar..212..180C}. If this attribution is correct, the sudden flattening of the 2004~EW$_{95}$ spectrum at $\lambda>0.8~\mu$m may potentially be the edge of a broad saponite absorption band centered in the $0.9-1.1~\mu$m region \citep[see][]{2011Icar..212..180C}. Without coverage of this region, however, we cannot rule out the possibility that our spectrum at $0.8-0.9~\mu$m is instead just continuum. The weaker band centered near $0.77~\mu$m is only tenuously detected and does not lend itself well to interpretation. Its center is just redward of the $0.70-0.75~\mu$m range where serpentine group phyllosilicate absorption bands are found \citep{2011Icar..212..180C}. The presence of a similar weak feature in the spectrum of 748 Simeisa \citep[Figure \ref{fig:ew95};][]{1989Sci...246..790V} indicates that it could be real, but without a firmer detection we choose not to overinterpret the data. Higher quality reflectance spectra, especially covering $\lambda>0.9~\mu$m, would aid future efforts to confirm the presence of aqueously altered material on 2004~EW$_{95}$ and interpret its mineralogy \citep[e.g.,][]{2011Icar..212..180C,2015aste.book...65R}.

\citet{2022PSJ.....3..153R} predict that if 2004~EW$_{95}$ is aqueously altered it should have a ``sharp type" spectral shape at $\lambda\sim3.0~\mu$m, defined by the presence of a $2.7~\mu$m OH band. This prediction is plausible, but it is worth noting that the presumed dynamical history and long term thermal evolution of 2004~EW$_{95}$ are not comparable to those of sharp type objects observed in the outer main asteroid belt. Any bound water and/or water ice present on the surface of 2004~EW$_{95}$ may act to broaden any potential $\sim3.0~\mu$m absorption band in its reflectance spectrum, as observed for Phoebe \citep[Saturn IX;][]{2005Natur.435...66C,2018AJ....156...23F}.

Comparison of our spectrum to coarse reflectance spectra derived from published colors of 2004~EW$_{95}$ reveals that some sets of single-epoch colors match the spectrum very well, while others do not (Figure \ref{fig:ew95}). The HST Cycle 17 colors reported by \citet{2015ApJ...804...31F} are in excellent agreement with our spectrum and, while their uncertainties are large, the $VRI$ colors from the same work also appear to be a good match. Within their uncertainties both epochs of $V-R$ colors reported by \citet{2013A&A...554A..49P} agree with the spectrum too, but the equivalent $V-I$ colors are wildly different (see Figure \ref{fig:ew95}). Because these $V-I$ colors disagree so strongly with all the other datasets, we consider them less reliable. The \citet{2015ApJ...804...31F} HST Cycle 18 colors more closely resemble the curvature observed in the X-Shooter spectrum of 2004~EW$_{95}$ reported by \citet{2018ApJ...855L..26S} than the bands observed in our new GMOS spectrum.

\begin{figure}
\centering
\includegraphics[scale=0.68]{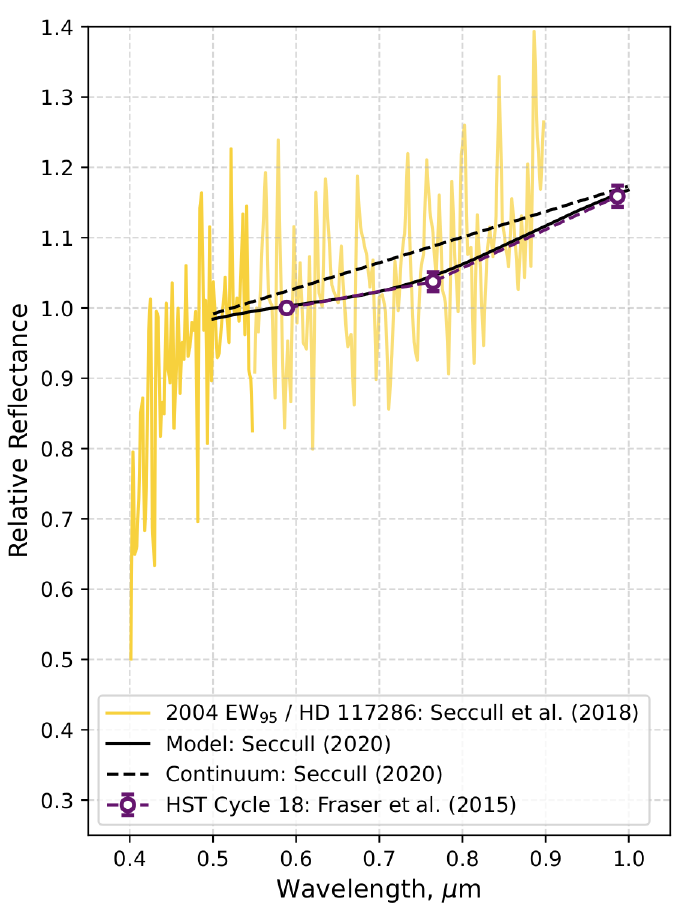}
\caption{The 2004~EW$_{95}$~/~HD~117286 reflectance spectrum published by \citet{2018ApJ...855L..26S} compared to the model of its $0.7~\mu$m band reported by \citet{2020PhDT........S}. We also compare the Cycle 18 HST colors reported for 2004~EW$_{95}$ by \citet{2015ApJ...804...31F}. All datasets are scaled to unit reflectance at $0.589~\mu$m.}\label{fig:ewmodel}
\end{figure}

To estimate the depth, width, central wavelength, and detection significance of their reported absorption band, \citet{2018ApJ...855L..26S} fitted a Gaussian model to their low S/N spectrum. An updated version of this model was created later \citep[see Chapter 6 of][where fitting methodology and parameter covariances are presented]{2020PhDT........S}. We have plotted the newer Gaussian model ($Depth=4\pm1\%$, $\lambda_c=0.73\pm0.04~\mu$m, $\sigma=0.12\pm0.04~\mu$m) over the spectrum from \citet{2018ApJ...855L..26S} in Figure \ref{fig:ewmodel}. While both the spectrum and the model have previously been reported, the excellent match between the model and the HST Cycle 18 photometry from \citet{2015ApJ...804...31F} has not. There is apparently consistency between the 2004~EW$_{95}$/HD~117286 reflectance spectrum and the HST Cycle 18 photometry; the same can be said for our new 2004~EW$_{95}$/HD~124523 reflectance spectrum and both the $VRI$ photometry and HST Cycle 17 photometry reported by \citet[][see also Figure \ref{fig:ew95}]{2015ApJ...804...31F}. The repeated observation of these two slightly different spectral shapes suggests either that there may be subtle variegation of the surface properties of 2004~EW$_{95}$ or that the various data reduction procedures used for each dataset may produce subtly inconsistent results. In either case, if the spectrum-color pairings that we have identified are real, the spectral variation they collectively represent appears to be small. \citet{2015ApJ...804...31F} reported no statistically significant spectral variation for 2004~EW$_{95}$ within the precision allowed by their collated photometric data. Unfortunately, the X-Shooter spectrum of 2004~EW$_{95}$ is too noisy to lend itself to a robust statistical comparison to our new GMOS spectrum. 

We note that to date all spectroscopic and photometric datasets capable of revealing signs of hydrated minerals on 2004~EW$_{95}$ do appear to suggest their presence. Assuming that observations of 2004~EW$_{95}$ have followed a stochastic sampling of rotational phase, which is more likely than repeat observations of a single phase, consistently observed signatures associated with hydrated minerals could indicate that 2004~EW$_{95}$ may be aqueously altered across much of its surface. Confirming this will, however, require a consistent, high S/N, and rotationally resolved spectroscopic survey of 2004~EW$_{95}$. 

\subsection{208996 (2003~AZ$_{84}$)}\label{sec:az}
\begin{figure*}
\centering
\includegraphics[scale=0.65]{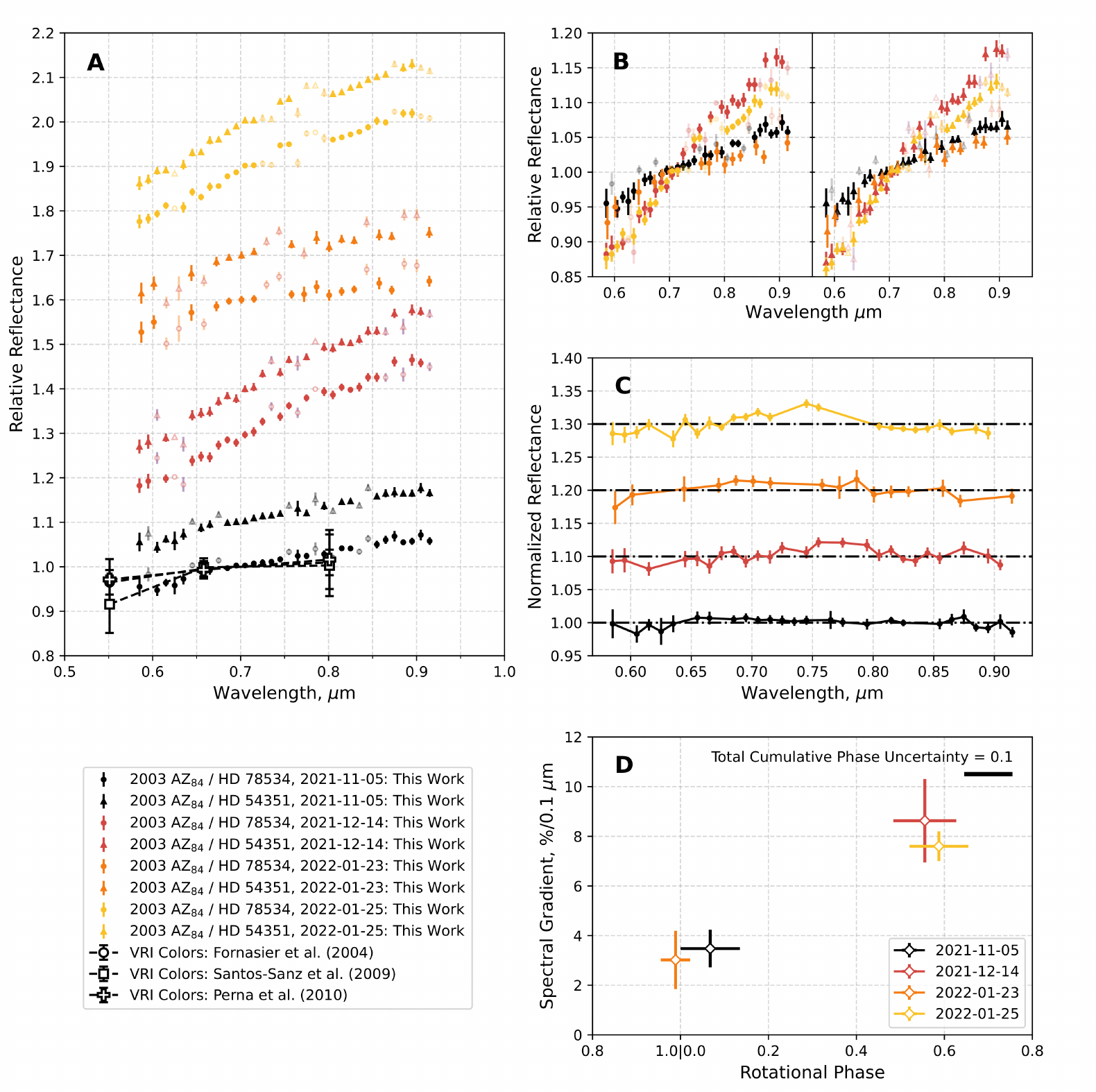}
\caption{The optical reflectance properties of 2003~AZ$_{84}$. \textbf{(A)}~Our new GMOS reflectance spectra of 2003~AZ$_{84}$ are presented alongside coarse reflectance spectra derived from single-epoch $VRI$ colors \citep{2004A&A...421..353F,2009A&A...494..693S,2010A&A...510A..53P}. Hollow points in the spectra are affected by incomplete background subtraction, telluric band residuals, proximity to the GMOS chip gaps, or a combination of these effects. All datasets are scaled to unit reflectance at $0.7~\mu$m and are vertically offset for clarity. Reflectance spectra observed at different epochs are offset in increments of +0.3, and those from the same epoch but calibrated with different stars are offset from each other by $\pm0.1$. \textbf{(B)}~The new spectra of 2003~AZ$_{84}$ are directly compared to each other to highlight the differences in their gradients. Spectra on the left are calibrated with HD~78534; those on the right are calibrated with HD~54351. \textbf{(C)}~Here we show our 2003~AZ$_{84}$ spectra following division by a line fitted to them across their full wavelength coverage. Spectra from each epoch are vertically offset for clarity in increments of +0.1. Only spectra calibrated with HD~78534 are shown. Points affected by residuals of telluric absorption or imperfect background subtraction have been omitted. \textbf{(D)}~A plot of the gradients (see Table \ref{tab:grads}) measured for each spectrum of 2003~AZ$_{84}$ calibrated with HD~78534 against the estimated rotational phase of the object at the time of observation. We adopt a nominal rotational period of $6.75\pm0.04$ hr \citep{2010A&A...522A..93T} and set a phase of zero to coincide with the start of our first observation of 2003~AZ$_{84}$. Rotational phase error bars show the rotational coverage of each observation. The black bar in the upper right corner shows the total rotational phase uncertainty accumulated over all four epochs; for each individual epoch the uncertainty is $<0.001$ (see text).\label{fig:az84}}
\end{figure*}

2003~AZ$_{84}$ is a large 3:2 resonant TNO (see Appendix \ref{sec:supp}). Clear detections of surface water ice have been reported for this object based on the presence of characteristic absorption bands in its reflectance spectrum at $1.5$ and $2.0~\mu$m \citep{2008AJ....135...55B,2009Icar..201..272G,2011Icar..214..297B}. It is predicted to be silicate-rich based on its infrared colors at $2.2<\lambda<5.0~\mu$m \citep{2021PSJ.....2...10F}, and sporadic observations of absorption bands at $\sim0.7~\mu$m have been attributed to hydrated minerals \citep{2004A&A...421..353F,2009A&A...508..457F,2008A&A...487..741A}. The $VRI$ colors of 2003~AZ$_{84}$ reported by \citet{2015A&A...577A..35P} do appear consistent with the potential presence of a hydrated mineral band when the photometric test from the Appendix \ref{sec:colors} is applied to them. Application of the test to the single-epoch colors of 2003~AZ$_{84}$ \citep{2004A&A...421..353F, 2009A&A...494..693S, 2010A&A...510A..53P} combined to make the average colors reported by \citet{2015A&A...577A..35P} suggest instead that a hydrated mineral band is not present in its reflectance spectrum (see Appendix \ref{sec:colors}).

The reportedly sporadic appearances of a $0.7~\mu$m hydrated mineral band in the spectrum of 2003~AZ$_{84}$ have led to predictions that its surface is variegated and that its spectrum is rotationally variable as a result. To maximize our odds of observing localized deposits of hydrated minerals, we chose to observe 2003~AZ$_{84}$ multiple times over the course of the 2021B semester. We did not set a specific cadence for our visits, however, and instead chose to let observations be performed as time and observing conditions allowed within the Gemini North observing queue.

All four of our new reflectance spectra are presented in Figure \ref{fig:az84} and the gradients we have measured from them are presented in Table \ref{tab:grads}. At all epochs the spectrum of 2003~AZ$_{84}$ was linear and showed no evidence of any absorption bands. While we can rule out the presence of detectable hydrated minerals on the surface of 2003~AZ$_{84}$ at the rotational phases where we observed it, we cannot rule out the possibility that localized deposits may have been previously observed at phases we did not cover \citep{2009A&A...508..457F}. Panel D in Figure \ref{fig:az84} shows the rotational phase coverage of our observations assuming a nominal rotational period of $6.75\pm0.04$~hr \citep{2010A&A...522A..93T}. The 0.04~hr rotational period uncertainty translates to a phasing uncertainty of 0.006 per rotation. Our observations targeting 2003~AZ$_{84}$ never took more than one hour to complete, so the maximum phasing uncertainty of each individual observing epoch is very small, at $<0.001$. By summing the phasing uncertainty in quadrature over the 287.7 rotations completed by 2003~AZ$_{84}$ between the start of our first observing epoch and the end of our last, we estimate that the total cumulative phase uncertainty across all four epochs is $\sim0.1$ (see Figure \ref{fig:az84}). 

Although no absorption bands were detected in the optical reflectance spectrum of 2003~AZ$_{84}$, we observed significant rotational variation in its gradient. At phases separated by half a rotation, the spectral gradient of 2003~AZ$_{84}$ was observed to vary by a factor $>2$, between $\sim3.5\%/0.1~\mu$m and $\sim8.5\%/0.1~\mu$m (see Figure \ref{fig:az84} and Table \ref{tab:grads}). Published spectra and colors favor the lower of these values \citep[e.g.,][]{2004A&A...421..353F,2009A&A...508..457F,2008A&A...487..741A,2009A&A...494..693S,2010A&A...510A..53P}. As continuum gradient in a spectrum can be sensitive to multiple confounding factors at the observation and data reduction stages, we performed multiple checks and tests to assess the veracity of the redder gradient measurement (details in Appendix \ref{sec:colortests}); ultimately, low level variations in the gradient could be induced by adjusting some data reduction parameters, but none could reproduce a variation of the magnitude that we observe. Our observing strategy was sufficiently consistent between epochs that variations in the spectral gradient are unlikely to stem from it. It seems, therefore, that the color variation is real and intrinsic to the surface of 2003~AZ$_{84}$. There are multiple factors to support this claim. The gradient consistently correlates with rotational phase over multiple observing epochs, and the two phase/color groups are separated in phase by 0.4-0.5 rotations, which is much larger than the total cumulative phase uncertainty of our combined observations. That we measure gradients consistent with previous reports at some epochs demonstrates that the variable color is unlikely to be caused by a measurement error.  

Variation of the spectrum of 2003~AZ$_{84}$ is well documented, from variable strength of the 1.5 and $2.0~\mu$m water ice bands and variable appearance of the $1.65~\mu$m crystalline water ice band \citep{2010Icar..208..945M,2011Icar..214..297B}, to sporadic appearances of a $\sim0.7~\mu$m hydrated mineral band \citep{2009A&A...508..457F}. Additionally, \citet{2009Icar..201..272G} and \citet{2011Icar..214..297B} both observed a potential feature at $\lambda\sim2.3~\mu$m in their spectra of 2003~AZ$_{84}$ that is clearly absent in the spectrum reported by \citet{2008AJ....135...55B}. \citet{2009Icar..201..272G} tentatively suggested that, if real, the $2.3~\mu$m band may be associated with methanol ice on the surface of 2003~AZ$_{84}$, while \citet{2011Icar..214..297B} flatly attribute it to noise in the data. Our observation of the variable optical color of 2003~AZ$_{84}$ is therefore not necessarily unexpected, but it does raise questions about why it has appeared now. Most published colors and spectra reported for 2003~AZ$_{84}$ were observed at least 14 years ago, and there is a high likelihood that our line of sight is misaligned with its equatorial plane \citep{2017AJ....154...22D}. The simplest explanation for the onset of color variation may therefore be that our viewing aspect of 2003~AZ$_{84}$ has changed enough that previously unobservable regions of its surface have now come into view.\footnote{The true anomaly of 2003~AZ$_{84}$ has changed by $\sim15^{\circ}$ since 2008.}

Alone, our disk-integrated optical reflectance spectra of 2003~AZ$_{84}$ offer little direct information about the size, origin, or nature of the reddened part of its surface, but we can engage in some cautious speculation. The strength of water ice absorption bands in the spectrum of 2003~AZ$_{84}$ is known to vary \citep{2010Icar..208..945M,2011Icar..214..297B}. The color variation we see might be driven by variation in the abundance and/or grain size of water ice across the surface of 2003~AZ$_{84}$. Areas of the surface with shorter optical path lengths through surface water ice may have reflectance properties dominated instead by optically red refractory opaques \citep[see also][]{2023Icar..39515492S}. The spectra of these areas may present both weaker water ice absorption bands and redder optical continuum colors than those observed for icier regions.

Reddened surfaces on TNOs may instead be attributed to the presence of refractory organics, but the localized presence of these on the surface of 2003~AZ$_{84}$ is difficult to explain given the apparent absence of volatile ice absorption bands in its reflectance spectrum. An impact of a red TNO with the surface of 2003~AZ$_{84}$ could potentially result in the localized deposition of red material, but it is not possible to determine whether this hypothesis is realistic until the size of the reddened region is constrained through higher resolution rotational monitoring of the surface color. A better constraint of the orientation of the rotational pole of 2003~AZ$_{84}$ would also be helpful in this effort \citep[e.g.,][]{2017AJ....154...22D}. Cryovolcanic delivery of internally produced red refractory organics to the surface is a different mechanism observed to have potentially caused localized surface reddening on the surface of Pluto \citep[e.g.,][]{1993GeCoA..57.3341S,2017SciA....3E2093K,2019Icar..330..155C,2021Icar..35613786C,2022NatCo..13.1542S}. Whether a similar mechanism could have caused localized reddening on 2003~AZ$_{84}$, a TNO with a diameter one-third that of Pluto, remains to be seen, given that the interior of 2003~AZ$_{84}$ likely experienced much less internal heating than larger cryovolcanically active TNOs \citep[see][]{2015Icar..246...21M}.  

\section{Discussion \& Conclusions}\label{sec:dis}
We have observed the optical reflectance spectra of three TNOs with the aim of detecting absorption features near $0.7~\mu$m associated with the presence of hydrated surface minerals. We obtained a clear detection of such features in the reflectance spectrum of 2004~EW$_{95}$, which supports previous reports that this object exhibits spectral behavior associated with the presence of aqueous alteration products \citep{2018ApJ...855L..26S}. Repeated and mostly consistent detections of these signatures suggest that the majority of the surface of 2004~EW$_{95}$ may be aqueously altered. By contrast, no absorption bands of any kind were observed in any of our four reflectance spectra of 2003~AZ$_{84}$. We cannot, however, rule out the possibility that previously reported detections of a $0.7~\mu$m band \citep{2004A&A...421..353F,2009A&A...508..457F,2008A&A...487..741A} could have been substantiated by observations performed at rotational phases that we missed. 2003~AZ$_{84}$ is well known to be spectrally variable \citep[e.g.,][]{2009A&A...508..457F,2010Icar..208..945M}, and for the first time we report the observation of significant rotational variation of its optical color. Because variegation of the surface of 2003~AZ$_{84}$ is likely, we predict that, if present, any detectable hydrated minerals are likely to be concentrated in localized deposits. Four new optical reflectance spectra of 2004~GV$_{9}$ reveal no sign of hydrated minerals on its surface. The reflectance properties of 2004~GV$_{9}$ appear to be very consistent at $\lambda<0.65~\mu$m, but show subtle variation at longer wavelengths.

\subsection{Aqueous Alteration in the Primordial Trans-Neptunian Disk}
For reasons discussed in Section \ref{sec:intro}, the detection of hydrated mineral absorption bands in the remotely sensed reflectance spectrum of a TNO is somewhat surprising; nevertheless, multiple such cases are reported in the literature \citep{2003AJ....125.1554L,2004A&A...416..791D,2004A&A...421..353F,2009A&A...508..457F,2008A&A...487..741A,2009A&A...501..777G,2018ApJ...855L..26S}. By considering heating mechanisms with the potential to drive aqueous alteration in TNOs, we aim to put such detections in the context of predicted pathways for TNO thermal evolution \citep[see also][]{2004A&A...416..791D}. 

Insolation, electrical induction, collisions, and radioactive decay of short-lived radionuclides are all mechanisms with the potential to heat the surface of a minor planet, but even in combination it is challenging for them to deliver enough heat to raise the surface temperatures of TNOs into the range required for aqueous alteration of silicates ($150<T\lesssim400~$K; see Section \ref{sec:intro}). 

\subsubsection{Insolation}
The weak insolation experienced by TNOs has likely never been able to raise their peak surface temperatures enough to enable aqueous alteration \citep[e.g.,][]{2002Icar..160..300C}. Even a low albedo TNO \citep[$p_V\sim0.04$; e.g.,][]{2014ApJ...793L...2L} orbiting the Sun at the inner boundary of the primordial trans-Neptunian disk \citep[$r_H\sim15$~au;][]{2007AJ....134.1790M} would never achieve a subsolar surface temperature much above 100~K \citep[see also][]{2011AJ....141..103G}.   

\subsubsection{Electrical Induction Heating}
Electrical induction heating is resistive heating of electrically conductive material within a minor planet caused by currents potentially induced by its motion through the strong magnetic field and solar wind of the Sun during its T~Tauri phase \citep[e.g.,][]{1970Ap&SS...7..446S,2015aste.book..553W}. It is possible that this mechanism may have heated some asteroids in the inner solar system in a narrow range of circumstances, but its efficiency is expected to be negligible for TNOs owing to their large heliocentric distances \citep{2013ApJ...776...89M}.

\subsubsection{Radiogenic Heating}
Radiogenic heating caused by the decay of short-lived radionuclides, especially \textsuperscript{26}Al \citep[half-life $\sim0.72~$Myr;][]{2009Icar..204..658C} and \textsuperscript{60}Fe \citep[half-life $\sim2.6~$Myr;][]{2015PhRvL.114d1101W}, is routinely considered to have been a primary driver of metamorphic processes within TNOs as they formed \citep[][and references therein]{2008ssbn.book..213M,2008ssbn.book..243C,2020tnss.book..183G}. The ability of internal radiogenic heating to facilitate the aqueous alteration of silicates on the surfaces of TNOs was likely negligible, however. The extent of radiogenic heating within a forming minor planet is inversely proportional to both its ice/silicate abundance ratio and its porosity \citep[e.g.,][]{2011A&A...529A..71G}. Being ice-rich and porous \citep[e.g,][]{2011Icar..214..297B,2012AJ....143..146B,2020tnss.book..201N}, TNOs probably formed with less radionuclide-bearing silicate material compared to asteroids of a similar size. More critical, however, is the timing and rate of formation of TNOs. To experience significant radiogenic heating, a TNO must form early enough relative to the half-lives of short-lived radionuclides to accrete enough of them before they decay \citep[typically within 2~Myr following the formation of calcium-aluminum-rich inclusions;][]{2003EM&P...92..359M,2006Icar..183..283M,2011A&A...529A..71G}. To continue raising its internal temperature, a TNO must also grow rapidly and shrink its surface/volume ratio enough that the rate of radiative heat loss at its surface is lower than the rate of radiogenic heat production \citep[e.g.,][]{2008ssbn.book..213M}. Thermal models of TNOs undergoing simultaneous accretion, insolation, and radiogenic heating demonstrate that, if a TNO forms rapidly enough, its interior can be raised to temperatures required for aqueous alteration to begin; they also show, however, that the silicates at the surfaces of all TNOs would have remained cold and unaltered to depths $>>10$~km \citep[e.g.,][]{2001AJ....121.2792D,2002Icar..160..300C,2002ESASP.500...29M,2003EM&P...92..359M,2006Icar..183..283M,2011A&A...529A..71G,2015Icar..246...21M}. Additional consideration of other internal exothermic processes, such as water ice crystallization and the serpentinization process itself, which might have initiated once prerequisite temperatures were reached, appear not to change this outcome \citep[e.g.,][]{2011A&A...529A..71G,2015Icar..246...21M}.

\subsubsection{Cryovolcanism}
Eutectic cryolavas partly comprised of liquid water have the potential to form both inside and at the surfaces of TNOs, where they may aqueously alter silicates that they contact \citep[e.g.,][]{1992Icar..100..556K}. Any hydrated minerals formed, however, would probably be neither soluble nor buoyant in the cryolava \citep{1992Icar..100..556K,2008ssbn.book..213M}. As a result, cryovolcanic delivery of hydrated minerals from the interior to the surface is unlikely. Aqueously altered silicates formed at the surface in this way are also likely to be overlaid by the cryovolcanic flows that created them, to depths that are optically thick. This supposition is nominally supported by observations of cryovolcanic features on Pluto performed by New Horizons, which did not reveal the presence of any associated surface hydrated minerals \citep{2019Icar..330..155C,2021Icar..35613786C,2022NatCo..13.1542S}. We note, however, that the ability of New Horizons to detect any hydrated minerals on Pluto was limited by both the low optical spectal resolution and lack of wavelength coverage redward of $2.5~\mu$m offered by its Ralph imager and imaging spectrometer \citep{2008SSRv..140..129R}. Given the aforementioned issues related to the buoyancy and solubility of hydrated minerals in cryolava, the existence of detectable cryovolcanically formed or deposited hydrated minerals on the surface of Pluto seems unlikely. It has yet to be formally ruled out, however.

\subsubsection{Collisions}
Collisions may drive the appearance of surface hydrated minerals on TNOs in multiple ways. First, cratering experiments show that impacts are likely to cause localized melting of surface water ice on TNOs if they have sufficiently high energy \citep{2008ssbn.book..195L}. Such melts may facilitate aqueous alteration of silicates at the impact site but, like cryolava, may also obscure any hydrated minerals they create. Second, collisions also have the potential to excavate hydrated minerals buried in the subsurface of a TNO. Finally, if they have any, impactors could deliver their internal hydrated minerals to the surfaces of objects that they collide with.

Based on the above considerations, it seems plausible that many TNOs may have hydrated minerals in their interiors. If they formed in the inner trans-Neptunian disk, their shorter formation timescales could have facilitated accretion of greater quantities of actively decaying radionuclides capable of driving aqueous alteration of silicates through interior radiogenic heating. We ultimately predict, however, that unless they formed in the inner solar system, the primary mechanisms by which TNOs gain hydrated minerals at their surfaces are delivery, excavation, or formation by impacts. 

Observations of the surface of Charon hint at a plausible link between impacts and the distribution of aqueous alteration products on TNOs. The disk-integrated NIR reflectance spectrum of Charon is dominated by the presence of characteristic $1.5~\mu$m and $2.0~\mu$m water ice absorption bands \citep{2000Icar..148..324B,2000Sci...287..107B,2001AJ....121.1163D,2007ApJ...663.1406C,2010Icar..210..930M,2017Icar..284..394H}, but accurate modeling of the continuum requires the addition of a dark material that absorbs at $\lambda>1.9~\mu$m and reflects neutrally at $\lambda<1.9~\mu$m. \citet{2000Icar..148..324B} reported that multiple OH-bearing materials, including phyllosilicates, could provide a reasonable, if imperfect, correction to their models of Charon's spectrum. Spatially resolved compositional mapping performed by New Horizons during its flyby of the Pluto system revealed that the dark absorber is present across much of Charon's surface, but its abundance appears correlated with underlying geological features, including some of Charon's impact craters \citep{2021psnh.book..433P}. The uneven distribution of the dark absorber across Charon led \citet{2021psnh.book..433P} to predict that it was most plausibly delivered to Charon by impactors. If hydrated minerals are indeed a component of this dark material, this prediction aligns well with our own regarding the delivery of aqueously altered material to the surfaces of TNOs. 

\subsection{Synthesis}
Most TNOs currently accessible to spectroscopic observation are large enough that they were likely still undergoing late stage formation at a time well beyond the half-lives of \textsuperscript{26}Al and \textsuperscript{60}Fe \citep[e.g.,][]{2012AJ....143...63K,2016A&A...592A..63D}. With small unaltered objects likely being more abundant in the disk, especially at larger heliocentric distances, accretion of hydrated minerals onto the surfaces of large TNOs would have been the exception, rather than the rule. For those forming more rapidly in the inner trans-Neptunian disk, however, accretion of aqueously altered material into their surface layers may have been more frequent. As a result, such objects may have gained localized deposits of hydrated minerals on their surfaces. Spectroscopic observations of these objects performed at random viewing aspects and rotational phases could potentially exhibit variation of the kind we observe in the reflectance spectrum of 2003~AZ$_{84}$ \citep{2009A&A...508..457F}. Of course, if the $0.7~\mu$m bands reported in the spectrum 2003~AZ$_{84}$ are produced by hydrated minerals, we should expect their appearance to coincide with that of the $2.7~\mu$m OH band. 

The creation of a globally hydrated TNO (possibly such as 2004~EW$_{95}$) is very challenging given the above considerations. Such an object would have to form rapidly in the inner part of the trans-Neptunian disk such that its interior, and those of the objects around it, would become aqueously altered. It would then have to experience evenly distributed impacts across its surface to gain surface hydrated minerals, while also avoiding collisions with unaltered objects that may partially blanket it with unaltered material or cause sufficient mixing of surface material that hydrated minerals become undetectable. The surface of such a TNO would also somehow have to become relatively depleted of refractory carbonaceous material that may otherwise obscure hydrated minerals. The relative consistency of the reflectance spectrum of 2004~EW$_{95}$ appears to suggest that its surface has not been significantly evolved by collisions; in contrast, Saturn's moon Phoebe has been heavily evolved by collisions and shows relatively strong spectral variation across its surface \citep[e.g.,][]{2002Icar..155..375B,2005Natur.435...66C,2006Icar..180..453V,2018AJ....156...23F}. 

A globally aqueously altered TNO could plausibly form as a chunk of the aqueously altered interior of a large disrupted parent body, but the efficiency of collisional disruption of TNOs is extremely low, especially for the large objects currently accessible to spectroscopic observation \citep[e.g.,][]{2022AJ....164..261A}. For example, the likelihood that a $\sim300$~km diameter 3:2 resonant TNO like 2004~EW$_{95}$ is the product of collisional disruption is $<<10^{-4}$ \citep{2022AJ....164..261A}.

We therefore agree with the previous interpretation of \citet{2018ApJ...855L..26S}; based on both its spectral similarity to carbonaceous asteroids and its particular dynamical properties (see Section \ref{sec:ew}), 2004~EW$_{95}$ appears to be an object that formed alongside the aqueously altered outer main-belt asteroids before being scattered to its present orbit as a result of the migrations of the giant planets. Initial publications based on TNO reflectance spectra observed by JWST program 2418 at $0.8-5.2~\mu$m report that a $4.26~\mu$m absorption band attributable to CO\textsubscript{2} ice appears in the spectra of many TNOs \citep[e.g.,][]{2023PSJ.....4..130B}. CO\textsubscript{2} ice is thermodynamically stable beyond Neptune's present orbit; at closer proximity to the Sun its volatility increases, and sublimation becomes efficient at $\sim12-15$~au \citep[e.g.,][]{1985A&A...142...31Y,1992acm..proc..545S,2016AJ....152...90W,2017RSPTA.37560247M}. It therefore seems reasonable to predict that if 2004~EW$_{95}$ did indeed form at heliocentric distances below $\sim12$~au, its surface should be depleted of CO\textsubscript{2} ice and the $4.26~\mu$m CO\textsubscript{2} absorption band should be absent from its reflectance spectrum.  We also predict that any TNO bearing detectable hydrated minerals across its entire surface plausibly shares a similar origin to 2004~EW$_{95}$, and could be an important tracer of the solar system's dynamical evolution.       

\begin{acknowledgements}
A portion of this work is based on observational data obtained with the Gemini North telescope, which is located adjacent to the summit of Maunakea within the Maunakea Science Reserve. We wish to recognize the cultural significance and reverence that the summit of Maunakea has always had within the indigenous Hawaiian community. We are fortunate to be able to observe the solar system from this unique mountain, and appreciate the opportunity to do so.

Thanks to the Gemini Observatory staff who performed (J. Berghuis, J. Chavez, J. Font-Serra, Z. Hartman, J.-E. Heo, A. Lopez, D. May, L. Magill, R. Ruiz, K. Silva, J. Turner) and queued our observations. We are grateful to John Blakeslee for awarding us director's time to observe 2004~EW$_{95}$, to Xiaoyu Zhang for assisting our access to the literature, and to Sandy Leggett for benevolently rescuing GS-2022A-Q-313 from the ITAC cutting room floor. 

This research is based on data obtained at the international Gemini Observatory and processed with Gemini IRAF. Gemini Observatory is a program of NSF's NOIRLab, and is managed by the Association of Universities for Research in Astronomy (AURA) under a cooperative agreement with the National Science Foundation on behalf of the Gemini Observatory partnership: the National Science Foundation (United States), National Research Council (Canada), Agencia Nacional de Investigaci\'on y Desarrollo (Chile), Ministerio de Ciencia, Tecnolog\'ia e Innovaci\'on (Argentina), Minist\'erio da Ci\^encia, Tecnologia, Inovações e Comunica\c{c}\~oes (Brazil), and Korea Astronomy and Space Science Institute (Republic of Korea). Our spectra were obtained under Gemini programs GS-2021A-DD-101, GN-2021B-Q-313, GS-2022A-Q-313, and GS-2023A-FT-105. During this project, T.S. received support through a Gemini Science Fellowship. T.H.P. acknowledges support through FONDECYT Regular project 1201016 and CONICYT project Basal FB210003.

Publication of this research was funded by NSF's NOIRLab and the National Research Council Canada.

This research made use of services provided by the NASA Jet Propulsion Laboratory Solar System Dynamics group, including ephemerides computed with Horizons (via the web interface) and orbital elements retrieved from the Small-Body Database (https://ssd.jpl.nasa.gov). We also benefited from bibliographic services offered by the NASA Astrophysics Data System (https://ui.adsabs.harvard.edu/). 
\end{acknowledgements}

%

\vspace{5mm}
\facility{Gemini Observatory: Gemini North (GMOS-N), Gemini South (GMOS-S)}


\software{Astropy \citep{2013A&A...558A..33A}, Astroscrappy \citep{2001PASP..113.1420V, 2018zndo...1482019M}, Gemini IRAF \citep{1986SPIE..627..733T,1993ASPC...52..173T,2012SASS...31..159G}, Matplotlib \citep{2007CSE..9..90H}, NumPy \citep{harris2020array}, SciPy \citep{2020SciPy-NMeth}.}



\appendix
\section{Searching for Candidate Aqueously Altered TNOs with $VRI$ Colors}\label{sec:colors}
Many TNOs have reported photometric colors, especially in the Johnson-Cousins $VRI$ filters, which cover the region where $0.7~\mu$m hydrated mineral features are found \citep[e.g.,][]{2012AA...546A.115H,2015A&A...577A..35P}. The depths and breadths typical of these absorption bands \citep[$depth<7\%$, $width\sim0.25~\mu$m; e.g.,][]{2014Icar..233..163F} makes them very difficult to detect with $VRI$ colors, however. The broad and overlapping bandpasses of these filters wash out any subtle changes in the curvature of the spectrum.

To identify potentially aqueously altered TNOs in the \citet{2015A&A...577A..35P} color catalog, we followed the example of \citet{2006Icar..180..453V} in using the inversed formulae of \citet{1995PhDT........21H} to convert $VRI$ colors to $vwx$ colors from the Eight Color Asteroid Survey \citep{1985Icar...61..355Z}. We then used the test described by \citet{1994Icar..111..456V} to test for the potential presence of a $0.7~\mu$m hydrated mineral band. To summarize, the color conversion formulae are
\begin{equation}
v-w=((V-R)-0.355)/0.836,
\end{equation}
\begin{equation}
v-x=((V-I)-0.695)/0.875,
\end{equation}
where the conversion factors in the removal of the solar spectral signature \citep{2006Icar..180..453V}. The resulting $vwx$ colors are then converted to reflectance, $R$. If
\begin{equation}
(R_w-((R_x-R_v)0.4984))/R_v<0.99,\label{eqn:vilas}
\end{equation}
then the colors of the minor planet suggest the presence of a $0.7~\mu$m absorption band in its reflectance spectrum \citep{1994Icar..111..456V}.

We find that this color test performs well in hinting at the presence of a $0.7~\mu$m band in the spectra of TNOs, provided that the $VRI$ colors used are reliable. Our application of the test to averaged TNO colors yielded mixed results. Averaged colors are combined from observations taken at various, and typically random, rotational phases; this means that the number of TNOs that test positive for a $0.7~\mu$m band in a catalog of averaged colors should be expected to be a lower limit. For a given object, we ultimately recommend performing this test for each of its reported sets of single-epoch $VRI$ colors in order to determine whether the result of the test is reproducible. In Table \ref{tab:test} we present the results of each test performed for each of the objects presented in this paper.  

\begin{table*}
\begin{center}
\caption{$0.7~\mu$m Band $VRI$ Color Test Results}
\label{tab:test}
\begin{tabular}{p{2.9cm}p{2.cm}p{6.5cm}}
\hline\hline
Target & Y & N \\[1pt]
\hline
90568 (2004~GV$_{9}$) & (1), $0.89\pm0.05$ & ~ \\[1pt]
120216 (2004~EW$_{95}$) & (2), $0.95\pm0.04$ & (3), $1.06\pm0.07$ and $1.09\pm0.06$ \\[1pt]
208996 (2003~AZ$_{84}$) & (1), $0.89\pm0.11$ & (4), $1.01\pm0.04$; (5), $1.05\pm0.10$; (6), $1.00\pm0.06$\\[1pt]
\hline
\hline
\end{tabular}\\[2pt]
\end{center}
\small{\textbf{Note.} This table shows whether previously published sets of $VRI$ colors predict the presence of a $0.7~\mu$m band in the spectra of the three TNOs presented in this work. Values presented here are the result of the left-hand side of equation \ref{eqn:vilas}; objects with values $<0.99$ test positive for the potential presence of a concavity in their reflectance spectrum centered near $0.7~\mu$m. Numbers in the Yes (Y) and No (N) columns correspond to the following references and are presented alongside their associated values for the left-hand side of formula \ref{eqn:vilas}: (1) \citet{2015A&A...577A..35P}; (2) \citet{2015ApJ...804...31F}; (3) \citet{2013A&A...554A..49P}; (4) \citet{2004A&A...421..353F}; (5) \citet{2009A&A...494..693S}; (6) \citet{2010A&A...510A..53P}.} \\
\end{table*}

\subsection{90568 (2004~GV$_{9}$)}
The average $VRI$ colors of 2004~GV$_{9}$ published by \citet{2015A&A...577A..35P} are combined from various sources, none of which observed a full set of $VRI$ colors within a single epoch \citep{2008AJ....136.1502R,2016AJ....152..210T}. This means that it is not possible to determine whether the positive result of the test is due to the presence of a $0.7~\mu$m feature or rotational variation of the spectral gradient of 2004~GV$_{9}$. Although our positive test result (Table \ref{tab:test}) was uncertain, we decided to go ahead with observing 2004~GV$_{9}$ because its optical reflectance spectrum had never been published, and because \citet{2021PSJ.....2...10F} predict it to be silicate-rich.

\subsection{120216 (2004~EW$_{95}$)}
We initially tested the averaged $VRI$ colors of 2004~EW$_{95}$ published by \citet{2015ApJ...804...31F}, and they test positive for a potential absorption near $0.7~\mu$m. The colors are a combination of those published by \citet{2013A&A...554A..49P}, and those measured from previously unpublished photometry. Closer inspection of the colors later revealed an error in the average, however, caused by a minor mistake where the $V-I$ colors from \citet{2013A&A...554A..49P} were taken as $R-I$ colors. As discussed in Section \ref{sec:ew} the $V-I$ colors from \citet{2013A&A...554A..49P} are very inconsistent with all the other colors and spectra reported for 2004~EW$_{95}$, making it difficult to trust them. When tested, the colors derived by \citet{2015ApJ...804...31F} from the unpublished photometry of \citet{2013A&A...554A..49P} test positive, and, although uncertain, they match to our spectrum very well. We therefore consider $(V-R)=0.38\pm0.04$ and $(R-I)=0.41\pm0.07$ to be the most reliable $VRI$ colors for 2004~EW$_{95}$.     

\subsection{208996 (2003~AZ$_{84}$)}
The average $VRI$ colors of 2003~AZ$_{84}$ published by \citet{2015A&A...577A..35P} test positive for the presence of absorption near $0.7~\mu$m, but comparison of the individual colors combined to make the average reveals a puzzling inconsistency. The average $V-R$ and $R-I$ colors are combined from multiple sources \citep{2004A&A...421..353F,2009A&A...494..693S,2010A&A...510A..53P}, all of which report that the $R-I$ color of 2003~AZ$_{84}$ is both similar to its $V-R$ color and slightly redder than the solar $R-I$ \citep[$(R-I)_\odot=0.345\pm0.004$][]{2012ApJ...752....5R}; the $V-I$ color published by \citet{2008AJ....136.1502R} is also consistent with being approximately solar. It is not clear, then, how the average $R-I$ became noticably redder than all of these. We find that all sets of single-epoch $VRI$ colors of 2003~AZ$_{84}$ from \citet{2004A&A...421..353F}, \citet{2009A&A...494..693S}, and \citet{2010A&A...510A..53P} test negative for the presence of absorption features at $0.7~\mu$m. 

\section{Supplementary Data}\label{sec:supp}
Tables \ref{tab:obslog}-\ref{tab:config} present, in order, the log, geometry, and instrument configurations for our observations. Table \ref{tab:supplemental} presents supplementary physical and dynamical information about our TNO targets.
\begin{table*}
\begin{center}
\caption{Observing Log}
\label{tab:obslog}
\begin{tabular}{p{4.4cm}p{4.7cm}p{2.2cm}p{1.3cm}p{2.cm}p{1.7cm}}
\hline\hline
Target & UT Observation Date $\vert$ Time & T\textsubscript{exp} (s) & N\textsubscript{exp} & Airmass & IQ (\arcsec) \\[1pt]
\hline
HD~124523 & 2021-03-13 $\vert$ 06:52:16--06:55:56 & 3.0 & 4 & 1.035-1.036 & 0.58-0.69  \\[1pt]
120216 (2004~EW$_{95}$) & 2021-03-13 $\vert$ 07:08:24--09:02:04 & 720.0 & 8 & 1.006-1.068 & 0.65-0.79 \\[1pt]
\hline
HD~54351 & 2021-11-05 $\vert$ 13:38:02--13:40:53 & 3.0 & 4 & 1.027-1.028 & 0.79-0.89 \\[1pt]
208996 (2003~AZ$_{84}$), \textit{Epoch 1} & 2021-11-05 $\vert$ 13:54:56--14:49:43 & 360.0 & 8 & 1.069-1.183 & 0.88-1.19 \\[1pt]
HD~78534 & 2021-11-05 $\vert$ 15:03:18--15:06:09 & 3.0 & 4 & 1.084-1.087 & 0.81-1.10 \\[1pt]
\hline
HD~54351 & 2021-12-14 $\vert$ 12:20:25--12:23:25 & 5.0 & 4 & 1.007-1.008 & 0.89-1.01 \\[1pt]
208996 (2003~AZ$_{84}$), \textit{Epoch 2} & 2021-12-14 $\vert$ 12:40:55--13:38:56 & 360.0 & 8 & 1.028-1.042 & 0.92-1.27 \\[1pt]
HD~78534 & 2021-12-14 $\vert$ 13:59:15--14:02:14 & 5.0 & 4 & 1.013-1.013 & 0.94-1.10 \\[1pt]
\hline
HD~54351 & 2022-01-23 $\vert$ 07:19:35--07:22:37 & 5.0 & 4 & 1.141-1.146 & 0.83-0.92 \\[1pt]
208996 (2003~AZ$_{84}$), \textit{Epoch 3} & 2022-01-23 $\vert$ 07:36:52--08:03:56 & 360.0 & 4 & 1.317-1.463 & 0.98-1.67 \\[1pt]
HD~78534 & 2022-01-23 $\vert$ 08:17:25--08:20:27 & 5.0 & 4 & 1.403-1.414 & 1.27-1.55 \\[1pt]
\hline
HD~54351 & 2022-01-25 $\vert$ 10:23:53--10:26:56 & 5.0 & 4 & 1.049-1.051 & 0.96-1.02 \\[1pt]
208996 (2003~AZ$_{84}$), \textit{Epoch 4} & 2022-01-25 $\vert$ 10:40:44--11:35:01 & 360.0 & 8 & 1.026-1.061 & 0.85-1.07 \\[1pt]
HD~78534 & 2022-01-25 $\vert$ 11:44:27--11:47:33 & 5.0 & 4 & 1.022-1.023 & 0.91-1.03 \\[1pt]
\hline
HD~124523 & 2022-03-07 $\vert$ 07:33:54--07:35:38 & 5.0 & 4 & 1.028-1.029 & 1.07-1.61 \\[1pt]
90568 (2004~GV$_{9}$), \textit{Epoch 1} & 2022-03-07 $\vert$ 08:09:16--08:23:23 & 60.0 & 8 & 1.021-1.033 & 1.14-1.37 \\[1pt]
HD~157750 & 2023-03-07 $\vert$ 08:32:59--08:34:29 & 2.0 & 4 & 1.194-1.198 & 0.95-1.34 \\[1pt]
\hline
HD~124523 & 2023-04-19 $\vert$ 08:10:21--08:11:49 & 5.0 & 4 & 1.260-1.262 & 1.08-1.11 \\[1pt]
90568 (2004~GV$_{9}$), \textit{Epoch 2} & 2023-04-19 $\vert$ 08:25:53--08:48:47 & 135.0 & 8 & 1.117-1.173 & 0.92-1.05 \\[1pt]
\hline
HD~124523 & 2023-04-20 $\vert$ 07:39:08--07:40:37 & 5.0 & 4 & 1.187-1.191 & 0.85-0.90 \\[1pt]
90568 (2004~GV$_{9}$), \textit{Epoch 3} & 2023-04-20 $\vert$ 07:56:57--08:19:52 & 135.0 & 8 & 1.072-1.114 & 0.87-0.95 \\[1pt]
\hline
HD~124523 & 2023-04-25 $\vert$ 04:42:54--04:45:25 & 5.0 & 4 & 1.019-1.025 & 0.64-0.75 \\[1pt]
90568 (2004~GV$_{9}$), \textit{Epoch 4} & 2023-04-25 $\vert$ 04:55:55--05:19:13 & 135.0 & 8 & 1.019-1.039 & 0.71-0.94 \\[1pt]
\hline
\end{tabular}\\[2pt]
\end{center}
\small{\textbf{Note.} For each target we present the UT observation date and time, the integration time per exposure (T\textsubscript{exp}), the number of exposures (N\textsubscript{exp}), and the airmass at which they were observed. The range of estimated IQ values presented for each target is the range of FWHMs measured from Moffat profiles \citep{1969A&A.....3..455M} fitted to the set of spatial profiles produced by median collapsing each of a target's reduced 2D spectroscopic exposures along the dispersion axis.} \\
\end{table*}

\begin{table*}
\begin{center}
\caption{Observation Geometry}
\label{tab:obsgeom}
\begin{tabular}{p{4.4cm}p{4.7cm}p{2.2cm}p{1.3cm}p{2.cm}p{1.7cm}}
\hline\hline
Target & R.A. (hr)~~~~~Decl. (deg) & r\textsubscript{H} (au) & $\Delta$ (au) & $\alpha$ (deg) & m\textsubscript{\textit{V}} (mag) \\[1pt]
\hline
120216 (2004~EW$_{95}$) & 15~23~45.58~~-34~58~08.4 & 27.045 & 26.624 & 1.93 & $21.3\pm0.4$  \\[1pt]
208996 (2003~AZ$_{84}$) \textit{Epoch 1} & 08~33~46.21~~+07~05~59.6 & 44.135 & 44.054 & 1.29 & $20.26\pm0.09$ \\[1pt]
208996 (2003~AZ$_{84}$) \textit{Epoch 2} & 08~32~39.68~~+06~59~36.4 & 44.123 & 43.450 & 0.95 & $20.21\pm0.09$ \\[1pt]
208996 (2003~AZ$_{84}$) \textit{Epoch 3} & 08~29~53.03~~+07~04~22.8 & 44.110 & 43.148 & 0.28 & $20.16\pm0.09$ \\[1pt]
208996 (2003~AZ$_{84}$) \textit{Epoch 4} & 08~29~42.92~~+07~04~55.0 & 44.109 & 43.145 & 0.26 & $20.16\pm0.09$ \\[1pt]
90568 (2004~GV$_{9}$) \textit{Epoch 1} & 15~27~28.76~~-25~28~36.5 & 39.762 & 39.404 & 1.34 & $20.4\pm0.3$ \\[1pt]
90568 (2004~GV$_{9}$) \textit{Epoch 2} & 15~31~45.32~~-25~11~09.3 & 39.824 & 38.939 & 0.70 & $20.3\pm0.2$ \\[1pt]
90568 (2004~GV$_{9}$) \textit{Epoch 3} & 15~31~40.71~~-25~10~52.8 & 39.824 & 38.931 & 0.68 & $20.3\pm0.2$ \\[1pt]
90568 (2004~GV$_{9}$) \textit{Epoch 4} & 15~31~17.31~~-25~09~27.1 & 39.825 & 38.897 & 0.57 & $20.3\pm0.2$ \\[1pt]
\hline
\end{tabular}\\[2pt]
\end{center}
\small{\textbf{Note.} For the median time of each TNO observation, we present the target's coordinates, heliocentric distance ($r$), geocentric distance ($\Delta$), phase angle ($\alpha$), and estimated apparent V-band magnitude ($m_V$). $m_V$ is calculated with $m_V = H_V + 5log_{10}({r_H}\Delta) + \beta\alpha$, where $H_V$ values were taken directly from the \textit{TNOs are Cool} public database \citep{2012A&A...541A..93M,2012A&A...541A..94V}. $\beta$ values for 2003~AZ$_{84}$ and 2004~GV$_{9}$ were taken directly from \citet{2008AJ....136.1502R}. $\beta=0.157\pm0.017$ was estimated for 2004~EW$_{95}$ by averaging $\beta$ values published by \citet{2007AJ....133...26R} for TNOs with $H_V>4$.} \\
\end{table*}

\begin{table*}
\begin{center}
\caption{Instrument Configuration}
\label{tab:config}
\begin{tabular}{p{4.4cm}p{4.7cm}p{2.2cm}p{1.3cm}p{2.cm}p{1.7cm}}
\hline\hline
Target & Instrument $\vert$ Program & Grating $\vert$ Filter & $\lambda_C$ ($\mu$m) & Slit Width (\arcsec) & $\lambda/\Delta\lambda$  \\[1pt]
\hline
120216 (2004~EW$_{95}$) & GMOS-S $\vert$ GS-2021A-DD-101 & R150 $\vert$ OG515 & 0.70 & 1.0 & $\sim315$ \\[1pt]
208996 (2003~AZ$_{84}$) & GMOS-N $\vert$ GN-2021B-Q-313 & R400 $\vert$ GG455 & 0.70 & 2.0 & $\sim480$  \\[1pt]
90568 (2004~GV$_{9}$) & GMOS-S $\vert$ GS-2022A-Q-313 & R150 $\vert$ GG455 & 0.56 & 1.0 & $\sim315$  \\[1pt]
90568 (2004~GV$_{9}$) & GMOS-S $\vert$ GS-2023A-FT-105 & R150 $\vert$ GG455 & 1.00 & 1.0 & $\sim315$  \\[1pt]
\hline
\end{tabular}\\[2pt]
\end{center}
\small{\textbf{Note.} We present the instrument configuration used to observe each TNO, including the instrument name and Gemini program number, the combination of grating and order blocking filter, the central wavelength setting for the grating ($\lambda_C$), the width of the slit used, and the approximate native resolving power of the spectrograph for each configuration at the blaze wavelength of the grating ($0.717~\mu$m for R150; $0.764~\mu$m for R400).} \\
\end{table*}

\begin{table*}
\begin{center}
\caption{Physical and Dynamical Properties of Minor Planets 90568, 120216, and 208996}
\label{tab:supplemental}
\begin{tabular}{cccccccccc}
\hline\hline
Target & $H_V$~(mag) & $D$~km & $p_V$ & Source & $q$ (au) & $Q$ (au) & $a$ (au)& e & i (deg)\\[1pt]
\hline
90568 (2004~GV$_{9}$) & $4.23\pm0.10$ & $680\pm34$ & $0.077^{+0.0084}_{-0.0077}$ & (1) & 38.74 & 44.82 & 41.78 & 0.0728 & 22.02\\[1pt]
120216 (2004~EW$_{95}$) & $6.69\pm0.35$ & $291^{+20.3}_{-25.9}$ & $0.044^{+0.021}_{-0.015}$ & (2) & 26.97 & 51.56 & 39.27 & 0.3131 & 29.33\\[1pt]
208996 (2003~AZ$_{84}$) & $3.74\pm0.08$ & $727^{+61.9}_{-66.5}$ & $0.107^{+0.023}_{-0.016}$ & (2) & 32.32 & 46.40 & 39.36 & 0.1788 & 13.56\\[1pt]
\hline
\hline
\end{tabular}\\[2pt]
\end{center}
\small{\textbf{Note.} Here we present selected physical and orbital parameters for each of our targets, including absolute $V$-band magnitude ($H_V$), diameter ($D$), $V$-band geometric albedo $p_V$, orbital perihelion distance ($q$), orbital aphelion distance ($Q$), orbital semimajor axis ($a$), orbital eccentricity ($e$), and orbital inclination ($i$). Sizes, albedos, and absolute magnitudes are referenced from (1) \citet{2012A&A...541A..94V} and (2) \citet{2012A&A...541A..93M}. Orbital parameters are taken from the Small-Body Database maintained by the Jet Propulsion Laboratory Solar System Dynamics group.} \\
\end{table*}

\section{Spectrum extraction with MOTES}\label{sec:motes}
MOTES is our Modular Optimized Tracer and Extractor of Spectra (\textit{T. Seccull \& D.A. Kiersz 2024, in preparation}). Like \citet{2018ApJ...855L..26S}, MOTES fits 1D Moffat profiles \citep{1969A&A.....3..455M} to the average spatial point-spread function (PSF) of each 2D spectrum within multiple adjacent dispersion bins along its wavelength axis to track its wavelength-dependent center position, FWHM, and overall PSF shape. The first iteration of this process is used to locate the background regions that can be used for sky subtraction. For the TNOs we defined the background to be anywhere outside of boundaries set at $\pm{3}\times{FWHM}$ along the spatial axis from the center of the spectrum's spatial profile. The spatial profiles of the much brighter calibrator star spectra had better defined wings, so we often had to adjust the sky boundaries and set them at greater distances from the spatial profile center. The pixels in the sky regions were used to estimate and subtract the contribution of the background. In all cases the sky subtraction was performed for each wavelength element in turn, after the background pixels had been sigma-clipped at $3\sigma$ to remove any remaining residual bad pixels or cosmic rays. The background in each 2D spectrum of 2003~AZ$_{84}$ and 2004~EW$_{95}$ was modeled and subtracted with a linear fit to improve subtraction of prominent sky emission lines at longer wavelengths and, in some cases, subtraction of faint diffuse background sources that drifted into the slit during our observations of each TNO. A simple median sky subtraction was sufficient for all observations of 2004~GV$_{9}$ and our solar calibrator stars.

After sky subtraction we localized the spectrum a second time for extraction. The boundaries of the extraction aperture were set at $\pm{2}\times{FWHM}$ from the spatial profile center for all targets and calibrators. For each dispersion element in turn, the local spatial profile was estimated by taking the nearest average Moffat profile and the estimated local profile center (both determined during localization) and aligning the former with the latter. This recentered Moffat profile was truncated by setting its values to zero for regions outside the extraction aperture and was then normalized by dividing it by its sum. An optimal extraction \citep[similar to that described by][]{1986PASP...98..609H} was then performed with this normalized spatial profile.

\section{Notes on Spectrum Binning}\label{sec:binning}
\begin{figure}
\centering
\includegraphics[scale=0.7]{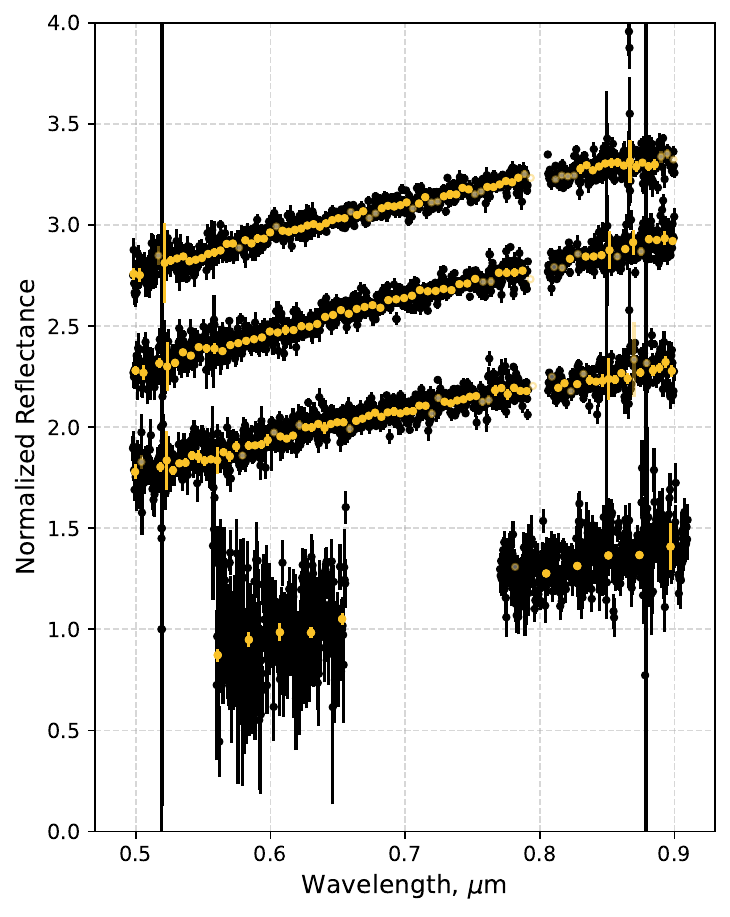}
\caption{Our GMOS-S 2004~GV$_{9}$/HD~124523 reflectance spectra presented as in Figure \ref{fig:gv9}, but with each spectrum presented in its binned (yellow) and unbinned (black) states. Spectra are ordered from bottom to top in order of observation date: 2022-03-07, 2023-04-19, 2023-04-20, and 2023-04-25. The lowest (first) binned spectrum has a binning factor of 60, and all other binned spectra are binned by a factor of 12. \label{fig:unbin_gv9}}
\end{figure}

\begin{figure}
\centering
\includegraphics[scale=0.7]{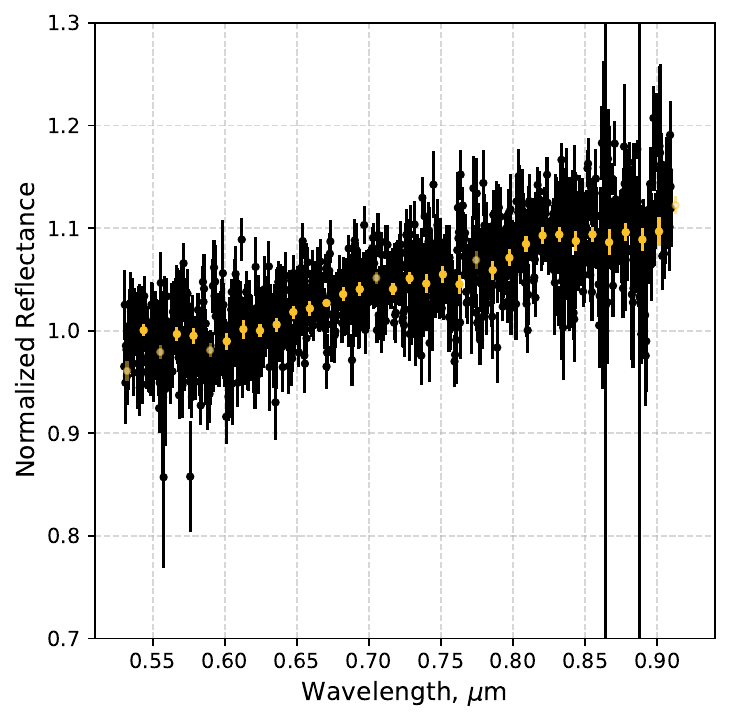}
\caption{Our new GMOS-S reflectance spectrum of 2004~EW$_{95}$ presented as in Figure \ref{fig:ew95}, but in both its binned ($\times30$; yellow) and unbinned (black) states.\label{fig:unbin_ew95}}
\end{figure}

\begin{figure}
\centering
\includegraphics[scale=0.7]{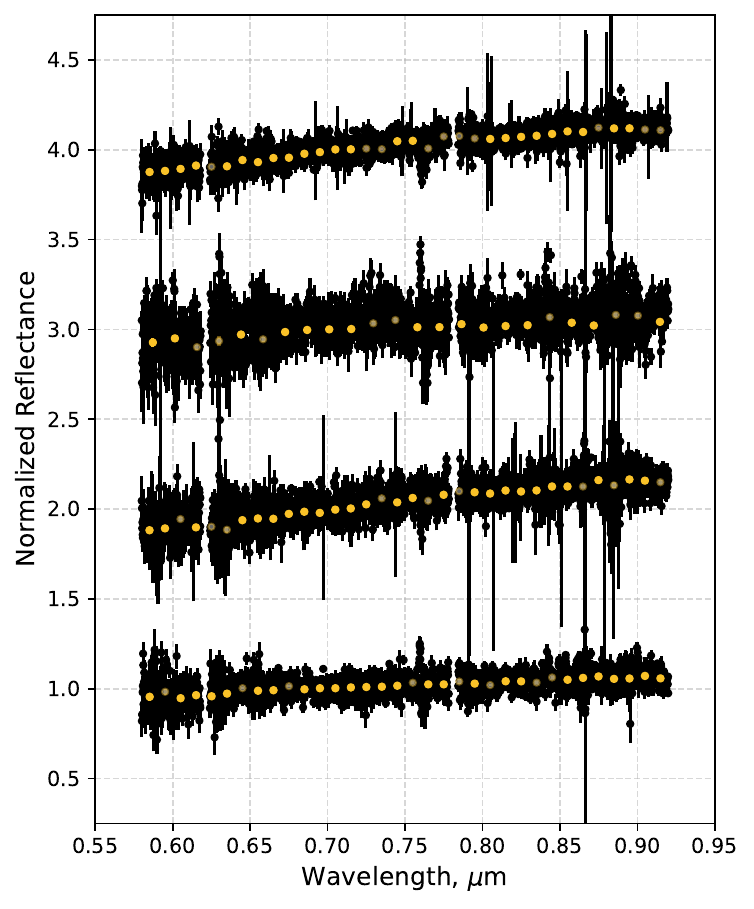}
\caption{Our GMOS-N 2003~AZ$_{84}$/HD~78534 reflectance spectra presented as in Figure \ref{fig:az84}, but with each spectrum presented in its binned (yellow) and unbinned (black) states. Spectra are ordered from bottom to top in order of observation date: 2021-11-05, 2021-12-14, 2022-01-23, and 2022-01-25. The binned spectrum from 2022-01-23 has a binning factor of 94; the binned spectra from all other epochs are binned by a factor of 66. \label{fig:unbin_az84}}
\end{figure}

As discussed in Section \ref{sec:obsred}, the bootstrapping method reported by \citet{2019AJ....157...88S} was used to bin our reflectance spectra. Here we elaborate on the selection process for the binning factor of each spectrum. For a given spectrum, we selected an appropriate binning factor through a process of iterative trial-and-error testing of different binning factors and comparative visual inspection of the binned spectrum, the unbinned 2D spectroscopic frames mapping the location of sky emission lines, and published maps of optical-NIR telluric absorption bands \citep{2015A&A...576A..77S}. 

As different binning factors are tested, the influence of residual sky emission lines, telluric absorption bands, and other instrumental effects are carefully identified and tracked. Suspected features of the spectrum, and absorption bands in particular, are also checked for consistent behavior (overall shape and location) at different binning factors to determine whether they are real (i.e., appearing consistently) or spurious. Through this method we aim to simultaneously achieve multiple things:

\begin{enumerate}
    \item Boost the S/N of the reflectance spectrum and clarify its intrinsic features. A related aim is the minimization of noisy residual structures in the data that are not intrinsic to the spectrum of the target (e.g., residual cosmic rays and sky emission lines).
    \item Preserve the real shape of the spectrum. Binning a spectrum may smear out weak features to the point that they become hidden; equally, some binning factors may amplify noisy residuals and produce features in the binned spectrum that do not really exist. 
    \item Minimize loss of spectral resolution.
    \item Maintain consistent spectral resolution between different spectra of a given target.
\end{enumerate}

Simultaneously meeting all these criteria for all spectra is often impossible, so we always give priority to aims one and two. In some cases, it therefore becomes necessary to use a large binning factor to beat down noisy residuals in a spectrum and adequately reveal its intrinsic shape. Fortunately, most solid state absorption bands expected to be present in the reflectance spectra of minor planets are very broad and can still be identified following a significant reduction in spectral resolution. In cases where satisfactory negation of residual features is impossible, we resort to flagging the data points they affect in the spectrum as bad.

For purposes of comparison, we respectively present our binned GMOS reflectance spectra from Figures \ref{fig:gv9}, \ref{fig:ew95}, and \ref{fig:az84} alongside their unbinned counterparts in Figures \ref{fig:unbin_gv9}-\ref{fig:unbin_az84}.

\section{Spectral Gradient Measurement}\label{sec:grads}
Our method of spectral gradient measurement is based on that described by \citet{2019AJ....157...88S}. For a binned reflectance spectrum, the set of values comprising each of its bins was bootstrapped 1000 times allowing for repeats and rebinned to produce 1000 resampled reflectance spectra with identical binning factor to the original. A linear regression was performed for each resampled spectrum over the required wavelength interval to measure the distribution of their spectral gradients. The final gradient measurement, $S'$, was defined as the mean of this distribution, and the standard error of the mean, $\sigma_{S'}$ (given by $\sigma/\sqrt{n-1}$, where $n=1000$), was taken as its uncertainty.

We estimate the final uncertainty of our gradient measurements to be the quadratic sum

\begin{equation}\label{eqn:suncertainty}
\delta{S'} = (\sigma_{S'}^2+(|S'_{O}-S'_{A}|)^2+(|A_{TNO}-A_{A*}|)^2)^{0.5},
\end{equation} 

where $\sigma_{S'}$ is the standard error of our measurement of $S'$, $S'_{O}$ is the gradient measured for the optimally extracted spectrum, $S'_{A}$ is the gradient measured for the aperture extracted spectrum, $A_{TNO}$ is the median airmass of the TNO observation, and $A_{*}$ is the median airmass of the calibrator star observation. \citet{2020ApJS..247...73M} showed that a spectral gradient uncertainty of $-0.92\%/~\mu$m is imparted to asteroid reflectance spectra per 0.1 unit airmass difference between the asteroid and its calibrator star when they are observed. We also account for this uncertainty, but conservatively set the value from \citet{2020ApJS..247...73M} to $\pm1.0\%/~\mu$m and divide it by 10 to convert it to our gradient units ($\%/0.1~\mu$m). Because our gradient uncertainty is $0.1\%/0.1~\mu$m per 0.1 unit airmass, we can substitute the airmass difference between TNO and calibrator star directly into equation \ref{eqn:suncertainty} in place of the gradient uncertainty derived from the airmass.

\section{Checking the Color Variation of 2003~AZ$_{84}$}\label{sec:colortests}
To test the veracity of our detection of variation in the color of 2003~AZ$_{84}$, we performed a number of checks and tests regarding our observation and data reduction procedures. Possible causes of slit losses at the time of observation were considered. We confirmed that the slit was correctly aligned to the average parallactic angle and used Gacq (the Gemini Observatory acquisition tool) to check how well all our targets were centered within the slit. Gacq showed that the targets were well centered, never being more than 2 pixels from the slit center (2 pixels is 0.8\% of the 2\arcsec~slit width). The calibrator stars were both solar twins \citep{2014A&A...563A..52P,2014A&A...572A..48R}, and are not expected to be significantly spectrally variable; even if they do vary, a synchronization of their variation, both with each other and with the rotation period of 2003~AZ$_{84}$, seems vanishingly unlikely. We varied the width of the extraction aperture when reducing our spectra and compared spectra extracted using both optimal extraction and aperture extraction. We found that an increase to the contribution of pixels further from the center of the spectrum's spatial profile, either by widening the extraction aperture or by removing the weightings imposed by optimal extraction (i.e., by just doing a simple aperture extraction), resulted in a modest reduction in the gradient of the final spectrum. This means that, at all epochs, using optimal extraction and a smaller aperture would limit the effects of imperfectly subtracted sky on the shape of the final spectrum. In all cases, however, changing the extraction parameters never changed the gradients of any of the spectra to the extent of the variation we observe in the data from epoch to epoch. Differences in the gradients measured for spectra extracted with different methods have been factored into the uncertainties of each of our gradient measurements. All flat-field frames were also checked and found to be consistent across all epochs to $<1\%$. As a result, we confirmed that the apparent color variation of 2003~AZ$_{84}$ is real. 
 

\bibliography{masterbib}{}
\bibliographystyle{aasjournal}



\end{document}